\documentclass[onecolumn,notitlepage,letterpaper,superscriptaddress,nofootinbib,longbibliography]{revtex4-2}

\usepackage[english]{babel}
\usepackage[T1]{fontenc}
\usepackage[utf8]{inputenc}
\usepackage{amsmath,accents}
\usepackage{amsfonts,color}
\usepackage{amssymb,float}
\usepackage{mathptmx}

\usepackage{graphicx}
\usepackage{subfigure}
\usepackage{color}

\usepackage[pdftoolbar = true, pdfstartview = FitH, pdfmenubar = true]{hyperref}
\hypersetup{
    colorlinks=true,
    linkcolor=red,
    filecolor=magenta,      
   citecolor=cyan
}

\newcommand{\lc}[1]{\accentset{\circ}{#1}}%Levi-Civita connection
\newcommand{\dd}{{\rm d}}%Operator d, i.e. non-italic

\newcommand{\be}{\begin{equation}}
\newcommand{\ee}{\end{equation}}
\newcommand{\bea}{\begin{eqnarray}}
\newcommand{\eea}{\end{eqnarray}}

\newcommand{\lptext}[1]{{\color{blue}{#1}}}

\begin{document}

\title{Stability of symmetric teleparallel scalar-tensor cosmologies with alternative connections}
  \author{Laur Järv} 
  \email{laur.jarv@ut.ee}
  \affiliation{Institute of Physics, University of Tartu, W.\ Ostwaldi 1, 50411 Tartu, Estonia}
  \author{Laxmipriya Pati}
  \email{lpriyapati1995@gmail.com}
  \affiliation{Institute of Physics, University of Tartu, W.\ Ostwaldi 1, 50411 Tartu, Estonia}
\begin{abstract}
    In symmetric teleparallel geometry the curvature and torsion tensors are assumed to vanish identically, while the dynamics of gravity is encoded by nonmetricity. Here the spatially homogeneous and isotropic connections that can accompany flat Friedmann-Lema\^itre-Robertson-Walker metric come in three sets. As the trivial set has received much attention, we focus on the two alternative sets which introduce an extra degree of freedom into the equations. Working in the context of symmetric teleparallel scalar-tensor gravity with generic nonminimal coupling and potential, we show that the extra free function in the connection can not play the role of dark matter nor dark energy, but it drastically alters the scalar field behavior. We determine the restrictions on the model functions which permit the standard cosmological scenario of successive radiation, dust matter, and scalar potential domination eras to be stable. However, the alternative connections also introduce a rather general possibility of the system meeting a singularity in finite time.
\end{abstract}
\maketitle
\section{Introduction}

The still unresolved origin of the observed accelerated expansion of the Universe known as dark energy, underlined by the significant tensions with the cosmic data the concordance $\Lambda$CDM model is evidently facing \cite{Perivolaropoulos:2021jda}, impels to look beyond Einstein's theory of general relativity (GR) \cite{Heisenberg:2018vsk,CANTATA:2021ktz,DiValentino:2021izs}. Perhaps the simplest and most widely studied type of GR extensions is to add a nonminimally coupled scalar field to the usual tensor degree of freedom. Such scalar-tensor models are remarkably successful in predicting the correct outcomes of early universe inflation \cite{Bezrukov:2007ep,Planck:2018jri}, but can also describe late universe dark energy \cite{Perrotta:1999am,Boisseau:2000pr} with effective ``phantom crossing'' behavior \cite{Perivolaropoulos:2005yv} and a possibility to address the cosmic tensions \cite{Ballardini:2020iws}. In the setup where freely falling test particles follow geodesics (the Jordan frame), nonminimal coupling manifests itself by making the Newtonian gravitational constant dependent on the value of the dynamical scalar field. However, large classes of scalar-tensor cosmologies exhibit an ``attractor mechanism'' whereby the scalar field spontaneously stabilizes since early matter dominated era \cite{Damour:1992kf}. This feature can explain how the theory passes various Solar system, astrophysical, and cosmological tests \cite{Will:2014kxa,Uzan:2010pm,Ooba:2016slp} and has been studied in detail in the literature \cite{Damour:1993id,Serna:1995pi,Mimoso:1998dn,Santiago:1998ae,Serna:2002fj,Jarv:2010zc,Jarv:2010xm,Jarv:2011sm,Jarv:2015kga,Dutta:2020uha}.

Remarkably, the Einstein-Hilbert lagrangian given by the Levi-Civita curvature scalar $\lc{R}$ is not the only way to derive the dynamics of general relativity \cite{BeltranJimenez:2019esp}. By invoking extra freedom in the connection characterized by torsion or nonmetricity, it is possible to demand that the geometry is teleparallel, i.e.\ the overall curvature tensor is zero. In the framework where nonmetricity also vanishes, one can then introduce (metric) teleparallel equivalent of general relativity (TEGR) \cite{Aldrovandi:2013wha,Krssak:2018ywd} by rewriting the Einstein-Hilbert lagrangian as $\lc{R}=-T+B_T$, where $T$ is the torsion scalar and $B_T$ is a boundary term that does not contribute to the field equations. Analogously, when torsion vanishes (and hence the affine connection is symmetric) we get symmetric teleparallel equivalent of general relativity (STEGR) \cite{Nester:1998mp} by writing $\lc{R}=Q+B_Q$ with nonmetricity scalar $Q$ and the respective boundary term $B_Q$. Until the matter lagrangian is left unchanged and still includes only coupling to the metric (or Levi-Civita connection), the  dynamics of TEGR and STEGR is equivalent to GR. The extra torsional and nonmetricity bits of connection completely drop out of the field equations, and remain arbitrary spectators with the only role to keep the curvature zero. However, the freedom to choose teleparallel connection arbitrarily may become restricted in the situations where the boundary term becomes relevant, like black hole energy or entropy \cite{BeltranJimenez:2021kpj,Gomes:2022vrc,Koivisto:2022oyt}.

With the same motivation that led to the extensions of GR, it is immediately tempting to introduce modified teleparallel lagrangians like $f(T)$ \cite{Bengochea:2008gz,Linder:2010py,Golovnev:2017dox,Hohmann:2017duq,BeltranJimenez:2018vdo} and $f(Q)$ \cite{BeltranJimenez:2017tkd,BeltranJimenez:2018vdo}, or extend the theory in the scalar-tensor manner by coupling a scalar field nonminimally to the torsion scalar \cite{Geng:2011aj,Hohmann:2018rwf} or nonmetricity scalar \cite{Jarv:2018bgs}. These extensions differ from their curvature based GR counterparts, which make them really interesting to study in search for new phenomenology \cite{Bahamonde:2021gfp}. In contrast to the TEGR and STEGR cases, the extra bits of connection are now endowed with their own equation. This raises an issue of how to proceed in solving the combined system of metric and connection equations, with different strategies conceivable \cite{Jarv:2019ctf}. One option that has been fruitful in several situations is to impose an ansatz for the connection which obeys the same set of spacetime symmetries as the metric \cite{Hohmann:2019nat}. For example in the case of cosmological Friedmann-Lema\^itre-Robertson-Walker (FLRW) spacetimes, the metric teleparallel connection (torsion only) comes in one family for spatial flatness and two families for spatial curvature, including no extra functions besides the scale factor already present in the metric \cite{Hohmann:2019nat,Hohmann:2020zre}. On the other hand, symmetric teleparallel connection (nonmetricity only) has three distinct options compatible with spatial flatness and one for spatial curvature, all cases endowed with an extra free function of time \cite{Hohmann:2021ast,DAmbrosio:2021pnd}. Different connections imply different field equations, and thus different dynamics.

The deeper theoretical discussions about extended teleparallel gravities in essence revolve around the issue of how much independent dynamics do the extra bits of connection really bring. The connection equation has only first order derivatives acting on the connection coefficients, and thus it looks more like a constraint equation. Furthermore, there is a gauge freedom in metric teleparallelism to perform a local Lorentz transformation that makes the spin connection to vanish (Weitzenböck gauge) usually at the expense of the tetrad assuming a more complicated form \cite{Krssak:2018ywd}, while in symmetric teleparallelism a general coordinate transformation can turn the affine connection to vanish (coincident gauge) typically making the metric more involved \cite{BeltranJimenez:2022azb,Bahamonde:2022zgj}. Thus the connection itself is a gauge degree of freedom, but it might conspire to hide more stuff in the tetrad or metric. A well respected procedure to determine the number of degrees of freedom in a theory is to perform Hamiltonian analysis, but so far there is no consensus on the results neither in $f(T)$ \cite{Li:2011rn,Ferraro:2018tpu,Blixt:2019mkt,Blagojevic:2020dyq,Blixt:2020ekl} nor $f(Q)$  \cite{Hu:2022anq,DAmbrosio:2023asf,Tomonari:2023wcs} gravity, except that there is something extra to the usual two tensor degrees of freedom of the metric. These extra degrees of freedom are hard to pinpoint precisely, though \cite{Ferraro:2018axk,Ferraro:2020tqk,Golovnev:2020zpv,BeltranJimenez:2021auj}. For instance, it is puzzling that no extra propagating modes show up among the linear perturbations around the cosmological backgrounds in $f(T)$ gravity \cite{Izumi:2012qj,Golovnev:2018wbh,Golovnev:2020aon,Bahamonde:2022ohm,Hu:2023juh}, while even the perturbations of the clearly independent scalar field get strangely blocked in simple scalar-torsion cosmology \cite{Golovnev:2018wbh,Raatikainen:2019qey,Ahmedov:2023num}. Seeking hints of the extra dynamical degrees of freedom one can go to higher perturbation orders \cite{BeltranJimenez:2020fvy}, or neglect the spacetime symmetry of the connection \cite{Golovnev:2020nln}.

In this context a closer look at the spatially flat FLRW connections in symmetric teleparallelism may offer new insights. The new function present in the connection could be taken as a clear instance of an extra degree of freedom. To be more precise, in one of the three classes this function completely decouples from the $f(Q)$ equations \cite{Hohmann:2021ast,DAmbrosio:2021pnd}, which reduce to the ones arising from trivial connection (that vanishes in Cartesian coordinates) \cite{BeltranJimenez:2017tkd}. This system actually coincides with the FLRW metric teleparallel equations \cite{Jarv:2018bgs} has been studied a lot \cite{Bahamonde:2021gfp}. However, in the two other classes the new function in the connection indeed manages to appear in the $f(Q)$ cosmological equations \cite{Hohmann:2021ast,DAmbrosio:2021pnd}. So far, there have been a few works about it, considering particular solutions and their properties \cite{Dimakis:2022rkd,Dimakis:2022wkj,Paliathanasis:2023ngs}, phase space features \cite{Shabani:2023nvm,Paliathanasis:2023nkb,Shabani:2023xfn}, and constraints from observational data \cite{Subramaniam:2023old,Shi:2023kvu}, all for certain specific models of $f(Q)$.

In this paper we investigate spatially flat FLRW cosmology in scalar-nonmetricity gravity, i.e.\ in the symmetric teleparallel analogue of the scalar-tensor theory \cite{Jarv:2018bgs}, utilizing the three classes of connections with respective symmetry \cite{Hohmann:2021ast,DAmbrosio:2021pnd}. We keep the nonminimal coupling function and potential completely generic, to cover as wide class of models as possible. The principal question is how does the extra function in the connection affect typical cosmological dynamics, from the radiation dominated epoch to matter domination, to the era of dark energy? Can it be itself a source of dark energy, or enable or disable the scalar field to behave as one? Can it allow phantom crossing for even minimally coupled scalar field?
Does the cosmological evolution spontaneously stabilize at some field values, i.e.\ is there an ``attractor mechanism'' like in the original scalar-tensor theory, or does the connection make the universe unstable? Finding answers to these questions can prepare ground for conducting direct assessments of selected more interesting types of models through comparisons with observational data. In the end we might even unravel a plausible explanation to the current conundrums in cosmology.

The manuscript is organized as follows. Section \ref{sec:action} recalls the geometrical setup of symmetric teleparallelism and the key features of scalar-nonmetricity gravity. All three classes of flat FLRW connections are explained in section \ref{sec:Spatially homogeneous and isotropic field configurations}. Section \ref{sec:scalartensorcosmology} presents the cosmological field equations arising for these connections, with attention on the dimensionality of the phase space and the possibility of emulating dark energy. Section \ref{sec: Limit of GR} introduces the expansion scheme for small perturbations, while section \ref{sec:Stability of standard cosmological regimes} explores in detail the stability of cosmological evolution in the dust matter dominated, radiation dominated, and scalar potential dominated regimes for the three classes connections. Finally, Sec. \ref{sec:conclusion} provides a discussion of the obtained results.
    
\section{Symmetric teleparallel gravities}\label{sec:action}

Symmetric teleparallel gravity assumes a geometric setup where the connection is characterized by identically vanishing curvature and torsion, while only nonmetricity is allowed to deviate from zero. The action can be constructed from the nonmetricity scalar which is equivalent to the Levi-Civita Ricci scalar up to a boundary term, thus providing a link to GR. 

\subsection{Geometric preliminaries} \label{subsec:Geometricpreliminaries}

A generic connection $\tilde{\Gamma}{}^{\lambda}_{\phantom{\alpha}\mu\nu}$ with 64 independent components can be decomposed into three parts \cite{Hehl:1976kj,Hehl:1994ue},
\begin{equation}
           \label{Connection decomposition}
                 \tilde{\Gamma}{}^{\lambda}_{\phantom{\alpha}\mu\nu} = 
                 \lc{\Gamma}^{\lambda}{}_{\mu\nu} +
                  K^{\lambda}_{\phantom{\alpha}\mu\nu}+
                  L^{\lambda}_{\phantom{\alpha}\mu\nu} \,,
\end{equation}
namely the Levi-Civita connection of the metric $g_{\mu\nu}$,
           \begin{equation}
          \label{LeviCivita}
                 \lc{\Gamma}^{\lambda}{}_{\mu \nu} \equiv \frac{1}{2} g^{\lambda \beta} \left( \partial_{\mu} g_{\beta\nu} + \partial_{\nu} g_{\beta\mu} - \partial_{\beta} g_{\mu\nu} \right) \,,
      \end{equation}
      contortion tensor
          \begin{equation}
          \label{Contortion}
                K^{\lambda}{}_{\mu\nu} \equiv \frac{1}{2} g^{\lambda \beta} \left( -T_{\mu\beta\nu}-T_{\nu\beta\mu} +T_{\beta\mu\nu} \right) =-\,K_{\mu}{}^{\lambda}{}_{\nu}\, ,
      \end{equation}
      and disformation tensor
     \begin{equation}
          \label{Disformation}
                L^{\lambda}{}_{\mu\nu} \equiv \frac{1}{2} g^{\lambda \beta} \left( -Q_{\mu \beta\nu}-Q_{\nu \beta\mu}+Q_{\beta \mu \nu} \right) = L^{\lambda}{}_{\nu\mu}  \,.
     \end{equation}
     Here contortion is built from torsion tensors that characterize the antisymmetric part of connection,
         \begin{equation}
          \label{TorsionTensor}
               T^{\lambda}{}_{\mu\nu}\equiv \tilde{\Gamma}{}^{\lambda}{}_{\mu\nu}-\tilde{\Gamma}{}^{\lambda}{}_{\nu\mu}\,,
      \end{equation}
       while disformation is constructed from nonmetricity tensors that are symmetric in the last two indices
         \begin{equation}
          \label{NonMetricityTensor}
               Q_{\rho \mu \nu} \equiv \nabla_{\rho} g_{\mu\nu} = \partial_\rho g_{\mu\nu} - \tilde{\Gamma}{}^\beta{}_{\mu \rho} g_{\beta \nu} -  \tilde{\Gamma}{}^\beta{}_{\nu \rho } g_{\mu \beta}  \,.
      \end{equation}
      These quantities along with curvature tensor
       \begin{eqnarray}
         \label{CurvatureTensor}
               R^{\sigma}\,_{\rho\mu\nu} &\equiv& \partial_{\mu}\tilde{\Gamma}{}^{\sigma}\,_{\nu\rho}-\partial_{\nu}\tilde{\Gamma}{}^{\sigma}\,_{\mu\rho}+\tilde{\Gamma}{}^{\sigma}\,_{\mu\lambda}\tilde{\Gamma}{}^{\lambda}\,_{\nu\rho}-\tilde{\Gamma}{}^{\sigma}\,_{\mu\lambda}\tilde{\Gamma}{}^{\lambda}\,_{\nu\rho}\,
         \end{eqnarray}
are the three key properties that characterize any connection. Zero curvature implies that the orientation of vectors does not change under parallel transport along a curve. Zero torsion implies that the connection is symmetric in the lower indices. Hence the imposition of vanishing curvature and torsion merits the name ``symmetric teleparallel'', and we denote it by ${\Gamma}{}^{\lambda}_{\phantom{\alpha}\mu\nu}$.

The connection ${\Gamma}{}^{\lambda}_{\phantom{\alpha}\mu\nu}$ has an interesting property that the scalar curvature of the Levi-Civita part of the connection, $\lc{R}$, can be expressed as 
        \begin{eqnarray}\label{R and Q}
              \lc{R}=Q+\lc{\nabla}_{\mu}(\hat{Q}^\mu-Q^\mu) \,,
       \end{eqnarray}
where the nonmetricity scalar and the two independent traces of the nonmetricity tensor are defined as\footnote{Note that some authors define $Q$ with the opposite overall sign, e.g.\ Ref.\ \cite{Hohmann:2021ast}.}
         \begin{align}\label{Qscalar}
              Q &\equiv -\,\frac{1}{4}\,Q_{\lambda\mu\nu}Q^{\lambda\mu\nu}+\frac{1}{2}\,Q_{\lambda\mu\nu}Q^{\mu\nu\lambda}+\frac{1}{4}\,Q_{\mu}Q^{\mu}-\frac{1}{2}\,Q_{\mu}\hat{Q}^{\mu}\,, \\
              Q_{\mu} &\equiv Q_{\mu\nu}\,^{\nu}\,, \qquad \qquad
              \hat{Q}_{\mu} \equiv Q_{\nu\mu}\,^{\nu}\,.
       \end{align}
It is also significant that the total Levi-Civita divergence part in \eqref{R and Q}, 
         \begin{align}
          \label{eq: B_Q}
             B_Q &=\lc{\nabla}_{\mu}(\hat{Q}^\mu-Q^\mu) 
        \end{align} 
         becomes a boundary term under a spacetime integral.

The idea behind symmetric teleparallel equivalent of general relatvity is the following. As the Einstein-Hilbert action of GR is given by the Levi-Civita curvature scalar $\lc{R}$, we can rewrite that action using the nonmetricity scalar $Q$ instead, and expect to keep the same dynamical content since the boundary term does not affect the field equations. Various extensions of the symmetric teleparallel theory like substituting $Q$ in the action by $f(Q)$ \cite{BeltranJimenez:2017tkd} or introducing a nonminimal coupling between $Q$ and a scalar field $\Phi$ \cite{Jarv:2018bgs}, however, lead to theories that are different from their counterparts $f(\lc{R})$ and scalar-tensor gravity originally formulated in the Riemannian geometry.

\subsection{Scalar-nonmetricity gravity} \label{subsec:Scanonmetricitygravity}
A symmetric teleparallel analogue of simple scalar-tensor action can be written as \cite{Jarv:2018bgs}
          \begin{equation}
            \label{Action}
                 S = \frac{1}{2\kappa^2} \int\mathrm{d}^4x \sqrt{-g} \left( \mathcal{A}(\Phi) Q -\mathcal B(\Phi) g^{\alpha\beta}\partial_\alpha\Phi \partial_\beta\Phi 
                 -2{\mathcal V}(\Phi)\right) + S_{\mathrm{m}}\,,
        \end{equation}
where $\kappa^2=8\pi G$. Like in the usual Riemannian scalar-tensor theory the nonminimal coupling function $\mathcal{A}$ sets the strength of the effective gravitational constant, $\mathcal{B}$ is the kinetic coupling function, and $\mathcal{V}$ is the scalar potential. We assume that the matter action $S_{\mathrm{m}}$ is the same as in GR, i.e.\ depending on the metric alone. 

With the help of introducing the so-called superpotential (or conjugate) tensor 
         \begin{eqnarray}
                P^\alpha{}_{\mu\nu}=-\,\frac{1}{4}Q^{\alpha}{}_{\mu\nu}+\frac{1}{2}Q_{(\mu}{}^\alpha{}_{\nu)}+\frac{1}{4}g_{\mu\nu}Q^\alpha-\frac{1}{4}(g_{\mu\nu}\hat{Q}^\alpha+\delta^\alpha{}_{(\mu}Q_{\nu)})\,
         \end{eqnarray}
as well as some geometric indentities, the field equations arising from the variation of the action \eqref{Action} with respect to the metric, symmetric teleparallel connection, and scalar field are the following \cite{Jarv:2018bgs,Bahamonde:2022esv}:
           \begin{subequations}
             \label{eq: scalar-tensor field equations}
           \begin{eqnarray}
                   \mathcal{A}(\Phi)\lc{G}_{\mu\nu}
                   +2\frac{\dd\mathcal{A}}{\dd\Phi} P^\lambda{}_{\mu\nu} \partial_\lambda \Phi + \frac{1}{2} g_{\mu\nu} \left( \mathcal{B}(\Phi) g^{\alpha\beta}\partial_\alpha\Phi \partial_{\beta} \Phi + 2 {\mathcal V}(\Phi)  \right) - \mathcal{B}(\Phi) \partial_{\mu} \Phi \partial_\nu \Phi &=&  \kappa^2\mathcal{T}_{\mu\nu} \,,
             \label{MetricFieldEqF}
                 \\
             \label{connectionEq}
                   \left( \frac{1}{2}Q_\beta + \nabla_\beta \right) \left[ \partial_\alpha \mathcal{A} \left( \frac{1}{2}Q_\mu g^{\alpha\beta}-\frac{1}{2}\delta^\alpha_\mu Q^\beta-Q_{\mu}\,^{\alpha\beta}+\delta^\alpha_\mu Q_{\gamma}\,^{\gamma\beta}\right)\right] &=&0\,,
                 \\
             \label{ScalarFieldEq}
                   2\mathcal B \lc{\square}\Phi 
                   +\frac{\dd\mathcal{B}}{\dd\Phi}g^{\alpha\beta}\partial_{\alpha}\Phi\partial_{\beta}\Phi + \frac{\dd\mathcal{A}}{\dd\Phi} Q -
                   2\frac{\dd\mathcal{V}}{\dd\Phi} &=&0 \,.
            \end{eqnarray}
         \end{subequations}
          Here $\lc{G}_{\mu\nu}$ is the Einstein tensor and $\lc{\square}$ the d'Alembert operator computed of the Levi-Civita part of the connection, while $\mathcal{T}_{\mu\nu}$ is the usual matter energy-momentum tensor. Taking the Levi-Civita covariant divergence of the metric field equations \eqref{eq: scalar-tensor field equations}, and using the connection equations \eqref{connectionEq} as well as the scalar field equation \eqref{ScalarFieldEq} gives the usual continuity equation of the matter fields \cite{Jarv:2018bgs}
          \begin{align}
              \label{eq: matter continuity equation}
                    \lc{\nabla}_{\mu} \mathcal{T}^{\mu}{}_{\nu} &= 0 \,.
          \end{align}
        It can be understood as a consequence of the matter part of the action only coupling to the metric (or Levi-Civita connection) whereby there are no matter (hypermomentum) sources in the connection equations \eqref{connectionEq}. More generally it is related to the diffeomorphims invariance of the matter action \cite{Koivisto:2005yk}.
         The same property is shared with the torsion based teleparallel scalar-tensor theory \cite{Hohmann:2018rwf}.
         Another essential point to highlight is that the connection equation \eqref{connectionEq} is not independent but intricately linked with the metric equations \eqref{MetricFieldEqF}. This interdependence arises from the Bianchi identity within symmetric teleparallelism, whereby by acting with the Levi-Civita covariant derivative on the metric equation one can arrive at the connection equation \cite{Heisenberg:2022mbo}. Consequently, when the metric equations are satisfied, the corresponding connection equations automatically follow. However, we will still keep the connection equations under view separately, as they express useful information about the fields.

When the scalar field is globally constant, Eq.\ \eqref{MetricFieldEqF} reduces to the Einstein's equation in GR with the value of the potential playing the role of the cosmological constant, while \eqref{connectionEq} and \eqref{ScalarFieldEq} immediately vanish. Therefore, the solutions of GR are trivially also the solutions of these scalar-tensor theories with the scalar field just being constant. If the nonminimal coupling function is fixed to unity, $\mathcal{A}(\Phi)\equiv 1$, and the kinetic and potential terms of the scalar field vanish, $\mathcal{B}(\Phi)\equiv\mathcal{V}(\Phi)\equiv 0$, the theory is reduced to a STEGR. If the nonminimal coupling function is unity, but the kinetic term of the scalar field is nontrivial, then we have a theory that is equivalent to a minimally coupled scalar field in GR. With the identifications $\mathcal{A}=f'(Q)$, $\mathcal{B}=0$, $2\mathcal{V}=Q f'(Q)-f(Q)$ the current model does also represent $f(Q)$ gravity \cite{Jarv:2018bgs}.

\section{Spatially homogeneous and isotropic field configurations} \label{sec:Spatially homogeneous and isotropic field configurations}

Spatially homogeneous and isotropic cosmological spacetimes are characterized by the Killing vectors of translations $\zeta_{T_i}$ and rotations $\zeta_{R_i}$, given in the spherical coordinates as 
\begin{subequations}
\label{eq: Killing FLRW}
\begin{align}
\zeta_{T_x}^\mu &= \begin{pmatrix} 0 & \chi \sin\theta \cos\phi & \frac{\chi}{r} \cos\theta \cos\phi & - \frac{\chi}{r} \frac{\sin\phi}{\sin\theta} \end{pmatrix}  \,, \\ 
\zeta_{T_y}^\mu &= \begin{pmatrix} 0 & \chi \sin\theta \sin\phi & \frac{\chi}{r} \cos\theta \sin\phi & \frac{\chi}{r} \frac{\cos\phi}{\sin\theta} \end{pmatrix}  \,, \\
\zeta_{T_z}^\mu &= \begin{pmatrix} 0 & \chi \cos\theta & - \frac{\chi}{r} \sin\theta & 0 \end{pmatrix}  \,, \\
\zeta_{R_x}^\mu &= \begin{pmatrix} 0 & 0 & \sin\phi & \frac{\cos{\phi}}{\tan{\theta}} \end{pmatrix} \,, \\
\zeta_{R_y}^\mu &= \begin{pmatrix} 0 & 0 & -\cos\phi & \frac{\sin\phi}{ \tan\theta} \end{pmatrix} \,, \\
\zeta_{R_z}^\mu &= \begin{pmatrix} 0 & 0 & 0 & -1 \end{pmatrix} \,,
\end{align}
\end{subequations}
where $\chi=\sqrt{1-k r^2}$ describes the curvature of the 3-space. In this paper we focus only upon the spatially flat case, thus $k=0$.
Since in the teleparallel context the connection is independent of the metric, imposing the symmetry fully means that the Lie derivatives of the metric and affine connection along these vectors vanish \cite{Hohmann:2019nat},
\begin{align}\label{LieD_mag}
\pounds_{\zeta}g_{\mu\nu}=0\,,\qquad
\pounds_{\zeta}{\Gamma}^{\lambda}\,_{\mu\nu}=0 \,.
\end{align}  
While it is well known that the metric which satisfies this condition is the Friedmann-Lema\^itre-Robertson-Walker, conveniently written as
\begin{align}
                \label{eq: FLRW metric}
                        ds^2 &= -dt^2 + a(t)^2 \left(dr^2 + r^2 d\theta^2 + r^2 \sin^2 \theta d\phi^{2} \right) \,,
\end{align}
the symmetric teleparallel connection components with the same spacetime symmetry were worked out only recently and practically simultaneously in Refs.\ \cite{Hohmann:2021ast} and \cite{DAmbrosio:2021pnd}. They come in three sets and are presented below by adopting the notation of Ref.\ \cite{Dimakis:2022rkd}. But before, let us note that matter energy-momentum, consistent with the cosmological symmetry is given by
          \begin{align}
                 \label{eq: FLRW matter}
                       \mathcal{T}_{\mu\nu} &= \left(\begin{matrix}\rho(t) & 0 & 0 & 0\\0 & a^{2}(t) p(t) & 0 & 0\\0 & 0 & r^{2} a^{2}(t) p(t) & 0\\0 & 0 & 0 & r^{2} a^{2}(t) p(t) \sin^{2}\theta \end{matrix}\right) 
              \end{align}
where we further assume a barotropic equation of state where the pressure is proportional to density, $p = \mathrm{w} \rho$. Similarly, the spatially homogeneous and isotropic scalar field can only depend on time,
             \begin{align}
                 \label{eq: FLRW scalar}
                      \Phi &= \Phi(t) \,.
             \end{align}
             
\subsection{Connection set 1}
The first set of spatially homogeneous and isotropic $k=0$ symmetric teleparallel connection can be presented as      
\begin{align}
               \label{eq: connection set 1}
                     \Gamma^\rho{}_{\mu\nu} &= \left[\begin{matrix}\left[\begin{matrix}\gamma(t) & 0 & 0 & 0\\0 & 0 & 0 & 0\\0 & 0 & 0 & 0\\0 & 0 & 0 & 0\end{matrix}\right] & \left[\begin{matrix}0 & 0 & 0 & 0\\0 & 0 & 0 & 0\\0 & 0 & - r & 0\\0 & 0 & 0 & - r \sin^{2}\theta\end{matrix}\right] & \left[\begin{matrix}0 & 0 & 0 & 0\\0 & 0 & \frac{1}{r} & 0\\0 & \frac{1}{r} & 0 & 0\\0 & 0 & 0 & - \sin\theta \cos\theta\end{matrix}\right] & \left[\begin{matrix}0 & 0 & 0 & 0\\0 & 0 & 0 & \frac{1}{r}\\0 & 0 & 0 & \cot\theta\\0 & \frac{1}{r} & \cot\theta & 0\end{matrix}\right]\end{matrix}\right]
\end{align}
where the four matrices in columns are labelled by the first index ${}^\rho$, and the entries of the matrices are specified by the last two indices ${}_{\mu \nu}$. This set was treated as case 1 ($K_2=K_3=0$) with $\gamma = K_1 = -K$ in \cite{Hohmann:2021ast} and as case $\Gamma^{(III)}_Q$ with $\gamma=C_1$ in \cite{DAmbrosio:2021pnd}. There are no extra restrictions on the function $\gamma(t)$. The nonmetricity scalar \eqref{Qscalar} computed from this connection is 
        \begin{align}
                 Q &= -6 H^2 \,.
            \end{align}

\subsection{Connection set 2}
The second set of spatially homogeneous and isotropic $k=0$ symmetric teleparallel connection can be presented as   
          \begin{align}
               \label{eq: connection set  2}
                     \Gamma^\rho{}_{\mu\nu} &= \left[\begin{matrix}\left[\begin{matrix}\gamma(t) + \frac{\dot{\gamma}(t)}{\gamma(t)} & 0 & 0 & 0\\0 & 0 & 0 & 0\\0 & 0 & 0 & 0\\0 & 0 & 0 & 0\end{matrix}\right] & \left[\begin{matrix}0 & \gamma(t) & 0 & 0\\\gamma(t) & 0 & 0 & 0\\0 & 0 & - r & 0\\0 & 0 & 0 & - r \sin^{2}\theta\end{matrix}\right] & \left[\begin{matrix}0 & 0 & \gamma(t) & 0\\0 & 0 & \frac{1}{r} & 0\\\gamma(t) & \frac{1}{r} & 0 & 0\\0 & 0 & 0 & - \sin\theta \cos\theta\end{matrix}\right] & \left[\begin{matrix}0 & 0 & 0 & \gamma(t)\\0 & 0 & 0 & \frac{1}{r}\\0 & 0 & 0 & \cot\theta\\\gamma(t) & \frac{1}{r} & \cot\theta & 0\end{matrix}\right]\end{matrix}\right]
          \end{align}
where by definition $\gamma(t) \neq 0$. This set was called case 3 ($K_2=0, K_3\neq 0$) with $\gamma = K_3 = K$ in \cite{Hohmann:2021ast} and case $\Gamma^{(I)}_Q$ with $\gamma=C_3$ in \cite{DAmbrosio:2021pnd}. The nonmetricity scalar characterizing the connection \eqref{eq: connection set  2} is 
       \begin{align}
                Q &= -6 H^2 + 9 H \gamma + 3 \dot{\gamma} \,.
        \end{align}

\subsection{Connection set 3}
The third set of spatially homogeneous and isotropic $k=0$ symmetric teleparallel connection can be presented as      
\begin{align}
           \label{eq: connection set 3}
                   \Gamma^\rho{}_{\mu\nu} &= \left[\begin{matrix}\left[\begin{matrix}- \frac{\dot{\gamma}(t)}{\gamma(t)} & 0 & 0 & 0\\0 & \gamma(t) & 0 & 0\\0 & 0 & r^{2} \gamma(t) & 0\\0 & 0 & 0 & r^{2} \gamma(t) \sin^{2}\theta\end{matrix}\right] & \left[\begin{matrix}0 & 0 & 0 & 0\\0 & 0 & 0 & 0\\0 & 0 & - r & 0\\0 & 0 & 0 & - r \sin^{2}\theta\end{matrix}\right] & \left[\begin{matrix}0 & 0 & 0 & 0\\0 & 0 & \frac{1}{r} & 0\\0 & \frac{1}{r} & 0 & 0\\0 & 0 & 0 & - \sin\theta \cos\theta\end{matrix}\right] & \left[\begin{matrix}0 & 0 & 0 & 0\\0 & 0 & 0 & \frac{1}{r}\\0 & 0 & 0 & \cot\theta\\0 & \frac{1}{r} & \cot\theta & 0\end{matrix}\right]\end{matrix}\right]
\end{align}
where by definition $\gamma(t) \neq 0$. This set was called case 2 ($K_2\neq0, K_3=0$) with $\gamma = K_2 = -a^2 K$ in \cite{Hohmann:2021ast} and case $\Gamma^{(II)}_Q$ with $\gamma=C_2$ in \cite{DAmbrosio:2021pnd}. The corresponding nonmetricity scalar is
        \begin{align}
                 Q &= -6 H^2 + 9 H \bar\gamma + 3 \dot{\bar\gamma}
       \end{align}
where $\bar\gamma = \frac{\gamma(t)}{a(t)^2}$.

%%%%%%%%%%%%%%%%%%%%%%%%%%%%%%%%%%%%%%%%%%%%%%%%%%%%%%%%%%%%
\section{Scalar-tensor cosmology} \label{sec:scalartensorcosmology}
      After these mathematical preliminaries, it is time to write out the cosmological equations, and ask whether the extra function $\gamma(t)$ in the connection does really constitute an independent degree of freedom, and whether it can mimic dark matter or dark energy in the description of our universe.
      
 \subsection{Connection set 1} \label{subsec:Connectionset1}
    Substituting the metric \eqref{eq: FLRW metric}, connection \eqref{eq: connection set 1}, matter \eqref{eq: FLRW matter} and scalar field \eqref{eq: FLRW scalar} into the field equations \eqref{eq: scalar-tensor field equations}, \eqref{eq: matter continuity equation} yields two nontrivial metric equations, scalar field equation, and matter continuity equation as follows:
       \begin{subequations}
        \label{eq.27}
           \begin{align}
        \label{eq: cosmology eq set 1 FR1}
                6 H^{2} \mathcal{A}(\Phi) - \dot{\Phi}^{2} \mathcal{B}(\Phi) - 2 \mathcal{V}(\Phi) &= 2 \kappa^{2} \rho \,, \\
                \label{eq: cosmology eq set 1 FR2}
                - 4 H \dot{\Phi} \mathcal{A}^{\, \prime}(\Phi) - \left(6 H^{2} + 4 \dot{H}\right) \mathcal{A}(\Phi) - \dot{\Phi}^{2} \mathcal{B}(\Phi) + 2 \mathcal{V}(\Phi) &= 2 \kappa^{2} \mathrm{w} \rho  \,, \\
        \label{eq: cosmology eq set 1 SE}
                -  6 H^{2} \mathcal{A}^{\,\prime}(\Phi) - ( 6 H \dot{\Phi} + 2 \ddot{\Phi} ) \mathcal{B}(\Phi) -  \dot{\Phi}^{2} \mathcal{B}^{\,\prime}(\Phi) - 2 \mathcal{V}^{\prime}(\Phi)&= 0  \,, \\
        \label{eq: cosmology eq set 1 ME}
               \dot{\rho} + 3 H (1 + \mathrm{w})\rho&= 0  \,.
        \end{align}
    \end{subequations}
As first pointed out in Ref.\ \cite{Jarv:2018bgs}, these equations coincide with the scalar-tensor cosmological equations in torsion based teleparallel gravity \cite{Hohmann:2018rwf}. They have been studied a lot in the teleparallel context \cite{Geng:2011aj,Wei:2011yr,Xu:2012jf,Gu:2012ww,Otalora:2013tba,Geng:2013uga,MohseniSadjadi:2013ywh,Kucukakca:2013mya,Skugoreva:2014ena,Jarv:2015odu,Skugoreva:2016bck,Jarv:2021ehj,Leon:2022oyy}, and we consider them here mainly for reference in comparison with the respective set 2 and 3 equations. In an analogous notation the corresponding scalar-tensor equations in the usual Riemannian theory can be found for instance in Ref.\ \cite{Jarv:2015kga}. The main difference here is that in the (symmetric) teleparallel case the scalar field equation \eqref{eq: cosmology eq set 1 SE} lacks matter sources. For minimally coupled ($\mathcal{A}^\prime=0$) theories the equations fully coincide.

Notice, that for the connection set 1 \eqref{eq: connection set 1} the connection equations \eqref{connectionEq} are satisfied identically, and $\gamma(t)$ is left completely arbitrary by the field equations. 
The remaining four equations above are not independent of each other, as we can take a time derivative of the Friedmann constraint \eqref{eq: cosmology eq set 1 FR1} and derive any of \eqref{eq: cosmology eq set 1 FR2}--\eqref{eq: cosmology eq set 1 ME} from the rest.
Furthermore, the Friedmann constraint \eqref{eq: cosmology eq set 1 FR1} gives an algebraic relation between the variables, making one them not independent. In the other words, the physical dynamics in the four-dimensional phase space of $\{\Phi, \dot{\Phi}, \rho, H\}$ takes place on a three-dimensional hypersurface determined by \eqref{eq: cosmology eq set 1 FR1}. In fact, we can explicitly reduce the system \eqref{eq.27} to a set of three first order differential equations that faithfully represent its dynamics. For instance we may eliminate $H$ and write       \begin{subequations}
          \label{eq: 1st order system set 1}
              \begin{align}
                    \dot{\Phi} =& \Pi \,, \\
                    \dot{\Pi} =& - \frac{\Pi^{2} \left(\mathcal{A}{\left(\Phi \right)} \mathcal{B}^\prime{\left(\Phi \right)} + \mathcal{B}{\left(\Phi \right)} \mathcal{A}^\prime {\left(\Phi \right)}\right)}{2 \mathcal{A}{\left(\Phi \right)} \mathcal{B}{\left(\Phi \right)}} - \frac{\kappa^{2} \rho \mathcal{A}^\prime {\left(\Phi \right)} \lptext{+} \mathcal{A}{\left(\Phi \right)} \mathcal{V}^\prime{\left(\Phi \right)}  \lptext{+}\mathcal{V}{\left(\Phi \right)} \mathcal{A}^\prime{\left(\Phi \right)}}{\mathcal{A}{\left(\Phi \right)} \mathcal{B}{\left(\Phi \right)}} \nonumber \\
                    & \mp \Pi \sqrt{\frac{3 (\Pi^{2} \mathcal{B}{\left(\Phi \right)} + 2 \kappa^{2} \rho + 2 \mathcal{V}{\left(\Phi \right)})}{2\mathcal{A}{\left(\Phi \right)}}} \,,\\
                    \dot{\rho} =& \mp  \left(1+ \mathrm{w}\right) \rho \sqrt{\frac{3 (\Pi^{2} \mathcal{B}{\left(\Phi \right)} + 2 \kappa^{2} \rho + 2 \mathcal{V}{\left(\Phi \right)})}{2\mathcal{A}{\left(\Phi \right)}}} 
              \end{align}
\end{subequations}
where the upper (lower) sign corresponds to the expanding $H>0$ (contracting $H<0$) branch of the Friedmann constraint \eqref{eq: cosmology eq set 1 FR1}. Any other combination of the variables like  $\{\Phi, \rho, H\}$ or  $\{\Phi, \tfrac{\dot{\Phi}}{H}, \tfrac{\rho}{H^2}\}$ when properly implemented would still have yielded a three-dimensional system. Below we will confirm that for the connection set 2 and 3 the phase space gets an additional independent dimension.
\subsection{Connection set 2} 
         \label{subsec: cosmo equations of set 2}
 From the metric \eqref{eq: FLRW metric}, connection set 2 \eqref{eq: connection set 2}, matter \eqref{eq: FLRW matter} and scalar field \eqref{eq: FLRW scalar} the two nontrivial metric equations \eqref{MetricFieldEqF}, one nontrivial connection equation \eqref{connectionEq}, the scalar field equation \eqref{ScalarFieldEq} and the matter continuity equation \eqref{eq: matter continuity equation} are
         \begin{subequations}
         \label{eq:cosmo equations set 2}
              \begin{align}
         \label{eq: cosmology eq set 2 FR1}
               3 \dot{\Phi} \gamma \mathcal{A}^{\,\prime}(\Phi) + 6 H^{2} \mathcal{A}(\Phi) - \dot{\Phi}^{2} \mathcal{B}(\Phi) - 2 \mathcal{V}(\Phi) &= 2 \kappa^{2} \rho \,, \\
         \label{eq: cosmology eq set 2 FR2}
               (3 \dot{\Phi} \gamma - 4 H \dot{\Phi} ) \mathcal{A}^{\,\prime}(\Phi) - \left(6 H^{2} + 4 \dot{H}\right) \mathcal{A}(\Phi) - \dot{\Phi}^{2} \mathcal{B} (\Phi) + 2 \mathcal{V}(\Phi) &= 2 \kappa^{2} \mathrm{w} \rho \,,  \\
         %\label{eq: cosmo connection equation set 2}
         \label{eq: cosmology eq set 2 CE}
                3 \gamma \left( \ddot{\Phi} \mathcal{A}^{\,\prime}(\Phi) + 3 H \dot{\Phi}  \mathcal{A}^{\,\prime}(\Phi) + \dot{\Phi}^{2} \mathcal{A}^{\, \prime \prime}(\Phi) \right) &=0 \,, \\
          \label{eq: cosmology eq set 2 SE}
                 ( - 6 H^{2} + 9 H \gamma + 3 \dot{\gamma} ) \mathcal{A}^{\,\prime}(\Phi) - (6 H \dot{\Phi} + 2 \ddot{\Phi} ) \mathcal{B}(\Phi) - \dot{\Phi}^{2} \mathcal{B}^{\,\prime}(\Phi)  - 2 \mathcal{V}^{\prime}(\Phi) &=0 \,, \\
          \label{eq: cosmology eq set 2 ME}
                 \dot{\rho} + 3 H (1 + \mathrm{w})\rho &= 0 \,.
              \end{align}
    \end{subequations}
    The nonzero function $\gamma$ appears in the equations always as multiplied by the derivatives of $\mathcal{A}$. Thus its has only effect in nonminimally coupled theories, i.e. extensions of STEGR. For the minimal couplings ($\mathcal{A}^\prime=\mathcal{A}^{\prime\prime}=0$) the function $\gamma$ fails to be present in the field equations, and remains completely arbitrary. In the case of minimal couplings the field equations coincide with those of set 1, and of the minimally coupled scalar field equations in general relativity. 

It is interesting to consider whether the connection contribution $\gamma$ in the equations can mimic dark matter or dark energy. For that it would need to have an available regime where it can act analogously to $\rho$ with $\mathrm{w}= 0$ or $-1$, respectively. This, however is not possible. If we compare the terms with $\gamma$ in Eqs.\ \eqref{eq: cosmology eq set 2 FR1} and \eqref{eq: cosmology eq set 2 FR2} to the terms with $\rho$, then the effective barotropic index we could assign to the $\gamma$ term would rather be $+1$, instead. Thus even by picking suitable model functions $\mathcal{A}$, $\mathcal{B}$, $\mathcal{V}$ that could allow to rewrite Eq.\ \eqref{eq: cosmology eq set 2 SE} as a respective effective continuity equation for $\gamma$, it can not contribute to effective dark matter or dark energy in the cosmological equations.

By construction, only four of the equations \eqref{eq:cosmo equations set 2} are independent, since any of them can be expressed as a linear combination of the time derivative of \eqref{eq: cosmology eq set 2 FR1} and the remaining equations. But the actual number of independent variables in the set $\{\Phi, \dot{\Phi}, \rho, H, \gamma \}$ is at first not so obvious, since the connection equation \eqref{eq: cosmology eq set 2 CE} does not provide dynamics for the independent connection function $\gamma$ but rather restrains the scalar field dynamics to
       \begin{align}
              \label{eq: cosmology eq set 2 Phi_tt}
                    \ddot{\Phi} &= -3 H \dot{\Phi} - \frac{\mathcal{A}^{\,\prime\prime}(\Phi)}{\mathcal{A}^{\,\prime}(\Phi)}\dot{\Phi}^2 \,.
         \end{align}
This could give an impression that although there is one extra variable and one extra equation, there is also one extra constraint that allows to eliminate $\ddot{\Phi}$ from the system, and reduce the number of independent variables by one. However, after eliminating $\ddot{\Phi}$, the Friedmann equation \eqref{eq: cosmology eq set 2 FR1} ceases to be a constraint that can lessen the number of independent variables, but rather assumes the role of a dynamical equation for $\Phi$. Thus in effect we are left with four first order equations for four independent variables $\{\Phi, \rho, H, \gamma \}$, and the phase space is four-dimensional. There is no way we can express any of these four in terms of the others.

Alternatively, we may treat \eqref{eq: cosmology eq set 2 Phi_tt} as a dynamical equation and use \eqref{eq: cosmology eq set 2 FR1} to eliminate $H$, writing the system \eqref{eq:cosmo equations set 2} as
      \begin{subequations}
        \label{eq: 1st order system set 2}
            \begin{align}
                 \dot{\Phi} =& \Pi \,, \\
        \label{eq: 1st order system set 2 Pi}
                 \dot{\Pi} =& - \frac{\Pi^{2}\mathcal{A}^{\prime\prime}(\Phi)}{\mathcal{A}^\prime(\Phi)}
                 \mp \Pi \sqrt{\frac{3 (\Pi^{2} \mathcal{B}{\left(\Phi \right)} -3 \Pi \gamma \mathcal{A}^\prime{\left(\Phi \right)} + 2 \kappa^{2} \rho + 2 \mathcal{V}{\left(\Phi \right)})}{2\mathcal{A}{\left(\Phi \right)}}} \,,\\
                 \dot{\rho} =& \mp  \left(1 + \mathrm{w}\right) \rho \sqrt{\frac{3 (\Pi^{2} \mathcal{B}{\left(\Phi \right)} -3 \Pi \gamma \mathcal{A}^\prime{\left(\Phi \right)} + 2 \kappa^{2} \rho + 2 \mathcal{V}{\left(\Phi \right)})}{2\mathcal{A}{\left(\Phi \right)}}} \,, \\
                 \dot{\gamma} =& - \frac{\Pi^{2} \left(2 \mathcal{A}{\left(\Phi \right)} \mathcal{B}{\left(\Phi \right)} \mathcal{A}^{\prime\prime}{\left(\Phi \right)} - \mathcal{A}{\left(\Phi \right)} \mathcal{A}^\prime{\left(\Phi \right)} \mathcal{B}^\prime{\left(\Phi \right)} - \mathcal{B}{\left(\Phi \right)} \left(\mathcal{A}^\prime{\left(\Phi \right)}\right)^{2}\right)}{3 \mathcal{A}{\left(\Phi \right)} \left(\mathcal{A}^\prime{\left(\Phi \right)}\right)^{2}} 
               - \frac{\Pi \gamma \mathcal{A}^\prime{\left(\Phi \right)}}{\mathcal{A}{\left(\Phi \right)}} \nonumber \\
                & + \frac{2 \left(\kappa^{2} \rho \mathcal{A}^\prime{\left(\Phi \right)} + \mathcal{A}{\left(\Phi \right)} \mathcal{V}^\prime{\left(\Phi \right)} + \mathcal{V}{\left(\Phi \right)} \mathcal{A}^\prime{\left(\Phi \right)}\right)}{3 \mathcal{A}{\left(\Phi \right)} \mathcal{A}^\prime{\left(\Phi \right)}} 
                \mp  \gamma \sqrt{\frac{3 (\Pi^{2} \mathcal{B}{\left(\Phi \right)} -3 \Pi \gamma \mathcal{A}^\prime{\left(\Phi \right)} + 2 \kappa^{2} \rho + 2 \mathcal{V}{\left(\Phi \right)})}{2\mathcal{A}{\left(\Phi \right)}}} \,,
          \end{align}
   \end{subequations}
which can be compared to the system \eqref{eq: 1st order system set 1} of set 1.
Again, there are four independent variables $\{\Phi, \dot{\Phi}, \rho, \gamma \}$, and for generic model functions it is not possible to reduce the system to a lower dimensional one.

As a surprise, according to Eq.\ \eqref{eq: cosmology eq set 2 Phi_tt} the kinetic coupling $\mathcal{B}(\Phi)$ and potential $\mathcal{V}(\Phi)$ play no direct role in the scalar field dynamics. Here for an expanding universe the first term on RHS dominates at small $\dot{\Phi}$ and acts as friction that slows down the scalar field evolution. The second term on RHS dominates at larger $\dot{\Phi}$ values and pushes $\dot{\Phi}$ to increasing or decreasing values, depending on the sign of $\tfrac{\mathcal{A}^{\,\prime\prime}(\Phi)}{\mathcal{A}^{\,\prime}(\Phi)}$. This means that for appropriate initial ``speed'' $\dot{\Phi}$ all solutions will reach a standstill at some arbitrary value of $\Phi$. However, for sufficiently large  initial ``speed'' the $\dot{\Phi}^2$ term can trigger ever stronger ``acceleration'' $\ddot{\Phi}$, throwing $\Phi$ towards infinity and causing a possible instability. If the coupling function  has an extremum ($\mathcal{A}^\prime=0$) at some value $\Phi_s$, then even small perturbations from static $\Phi$ can launch such unstable behavior. In the phase space the scalar field can not evolve past the value of $\Phi_s$ where $\tfrac{\mathcal{A}^{\,\prime\prime}}{\mathcal{A}^{\,\prime}}$ becomes singular, since depending on the initial conditions it is either forced to stop at $\Phi_s$ or meets a sudden singularity with $|\dot{\Phi}| \to \infty$ in finite time. 
    Thus by simply analyzing the field equations, we should become apprehensive about the stability of the solutions with the connection set 2.

\subsection{Cosmology of connection set 3} \label{subsec:Cosmology of connection set 3}
Finally, inserting the metric \eqref{eq: FLRW metric}, connection set 3 \eqref{eq: connection set 3}, matter \eqref{eq: FLRW matter} and scalar field \eqref{eq: FLRW scalar} to the field equations yields the two nontrivial metric equations \eqref{MetricFieldEqF}, one nontrivial connection equation \eqref{connectionEq}, the scalar field equation \eqref{ScalarFieldEq} and the matter continuity equation \eqref{eq: matter continuity equation} as
     \begin{subequations}
       \label{eq:cosmo equations set 3}
          \begin{align}
       \label{eq: cosmology eq set 3 FR1}
                6 H^{2} \mathcal{A}(\Phi) - 3 \bar\gamma \dot{\Phi} \mathcal{A}^{\,\prime}(\Phi) - \dot{\Phi}^{2} \mathcal{B}(\Phi) - 2 \mathcal{V}(\Phi) &= 2\kappa^{2} \rho  \\
      \label{eq: cosmology eq set 3 FR2}
               (\bar\gamma \, \dot{\Phi} - 4 H \dot{\Phi} ) \mathcal{A}^{\,\prime}(\Phi) - \left(6 H^{2} + 4 \dot{H}\right) \mathcal{A}(\Phi) - \dot{\Phi}^{2} \mathcal{B}(\Phi) + 2 \mathcal{V}(\Phi) &= 2 \kappa^{2} \mathrm{w} \rho  \\
     \label{eq: cosmology eq set 3 CE}
              - 6 \dot{\bar\gamma} \, \dot{\Phi} \mathcal{A}^{\,\prime}(\Phi) - 3 \bar{\gamma} \left( \ddot{\Phi} \mathcal{A}^{\,\prime}(\Phi) + 5 H \, \dot{\Phi} \, \mathcal{A}^{\,\prime}(\Phi) + \dot{\Phi}^{2} \mathcal{A}^{\,\prime\prime}(\Phi) \right)   &=0 \\
     \label{eq: cosmology eq set 3 SE}
             (- 6 H^{2} + 9 H \bar\gamma + 3 \dot{\bar\gamma} ) \mathcal{A}^{\,\prime}(\Phi) - (6 H \, \dot{\Phi}+ 2 \ddot{\Phi} ) \mathcal{B}(\Phi) -  \dot{\Phi}^{2} \mathcal{B}^{\,\prime}(\Phi) -  2 \mathcal{V}^{\prime}(\Phi) &=0 \\
     \label{eq: cosmology eq set 3 ME}
            \dot{\rho} + 3 H (1 + \mathrm{w}) \rho &= 0
         \end{align}
  \end{subequations}
   Like in the previous case of set 2, the nonzero function $\bar\gamma$ appears in the equations always as multiplied by the derivatives of $\mathcal{A}$, and thus has only effect in nonminimal theories. For the minimal coupling the function $\bar\gamma$ does not enter the field equations and remains completely arbitrary, while the equations themselves coincide with those of the minimally coupled scalar field equations in general relativity. 

Compared to the set 2 case, the $\bar\gamma$ term in \eqref{eq: cosmology eq set 3 FR1} comes with an opposite sign. Despite that, $\bar\gamma$ can not play the role of an effective dark matter or dark energy, since the effective barotropic index we could assign to the $\bar\gamma$ terms in \eqref{eq: cosmology eq set 3 FR1}--\eqref{eq: cosmology eq set 3 FR2} is $-\tfrac{1}{3}$, i.e.\ analogous to the spatial curvature term. In the remaining equations the $\bar\gamma$ terms do not quite act like the spatial curvature for generic model functions, but perhaps some combinations of $\mathcal{A}(\Phi)$, $\mathcal{B}(\Phi)$, $\mathcal{V}(\Phi)$ may indeed allow it to mimic such behavior even exactly.

    Again, by construction, only four of the equations \eqref{eq:cosmo equations set 3} are independent, since any of them can be expressed as a linear combination of the time derivative of \eqref{eq: cosmology eq set 3 FR1} and the remaining equations. The connection equation \eqref{eq: cosmology eq set 3 CE} contains $\ddot{\Phi}$ and $\bar\gamma$, and can be used to eliminate either of them, eventually leading to $\{\Phi, \rho, H, \gamma \}$ or $\{\Phi, \dot{\Phi}, \rho, H \}$ as independent variables.
For example, we may express from \eqref{eq: cosmology eq set 3 CE} 
     \begin{align}
    \label{eq: cosmology eq set 3 Phi_tt}
        \ddot{\Phi}=-\left(\frac{2\dot{\bar\gamma}}{\bar\gamma}+5H\right)\dot{\Phi}-\frac{\mathcal{A}^{\, \prime\prime}(\Phi)}{\mathcal{A}^{\,\prime}(\Phi)}\dot{\Phi}^{2}
    \end{align}
and substitute it in. 

Alternatively, we may use \eqref{eq: cosmology eq set 3 FR1} to eliminate $H$, and write the system \eqref{eq:cosmo equations set 3} as
   \begin{subequations}
       \label{eq: 1st order system set 3}
            \begin{align}
                \dot{\Phi} =& \Pi \,, \\
      \label{eq: 1st order system set 3 Pi}
                 \dot{\Pi} =& - \frac{\Pi^{2}\mathcal{A}^{\prime\prime}(\Phi)}{\mathcal{A}^\prime(\Phi)}
                -\frac{2 \Pi \dot{\bar\gamma}}{\bar\gamma}
                \mp 5 \Pi \sqrt{\frac{\Pi^{2} \mathcal{B}{\left(\Phi \right)} +3 \Pi \bar\gamma \mathcal{A}^\prime{\left(\Phi \right)} + 2 \kappa^{2} \rho + 2 \mathcal{V}{\left(\Phi \right)}}{6\mathcal{A}{\left(\Phi \right)}}} \,,\\
               \dot{\rho} =& \mp  \left(1 + \mathrm{w}\right) \rho \sqrt{\frac{3 (\Pi^{2} \mathcal{B}{\left(\Phi \right)} +3 \Pi \bar\gamma \mathcal{A}^\prime{\left(\Phi \right)} + 2 \kappa^{2} \rho + 2 \mathcal{V}{\left(\Phi \right)})}{2\mathcal{A}{\left(\Phi \right)}}} \,, \\
    \label{eq: 1st order system set 3 gamma}
               \dot{\bar\gamma} =& - \frac{\Pi^{2} \bar\gamma \left(2 \mathcal{A}{\left(\Phi \right)} \mathcal{B}{\left(\Phi \right)} \mathcal{A}^{\prime\prime}{\left(\Phi \right)} - \mathcal{A}{\left(\Phi \right)} \mathcal{A}^\prime{\left(\Phi \right)} \mathcal{B}^\prime{\left(\Phi \right)} - \mathcal{B}{\left(\Phi \right)} \left(\mathcal{A}^\prime{\left(\Phi \right)}\right)^{2}\right)}{\left(4 \Pi \mathcal{B}(\Phi) + 3 \bar\gamma \mathcal{A}^\prime{\left(\Phi \right)}\right) \mathcal{A}{\left(\Phi \right)} \left(\mathcal{A}^\prime{\left(\Phi \right)}\right)} 
               + \frac{3 \Pi \bar\gamma^2 \left(\mathcal{A}^\prime{\left(\Phi \right)}\right)^2}{\left(4 \Pi \mathcal{B}(\Phi) + 3 \bar\gamma \mathcal{A}^\prime{\left(\Phi \right)}\right) \mathcal{A}{\left(\Phi \right)} } \nonumber \\
               & + \frac{2 \bar\gamma \left(\kappa^{2} \rho \mathcal{A}^\prime{\left(\Phi \right)} + \mathcal{A}{\left(\Phi \right)} \mathcal{V}^\prime{\left(\Phi \right)} + \mathcal{V}{\left(\Phi \right)} \mathcal{A}^\prime{\left(\Phi \right)}\right)}{\left(4 \Pi \mathcal{B}(\Phi) + 3 \bar\gamma \mathcal{A}^\prime{\left(\Phi \right)}\right) \mathcal{A}{\left(\Phi \right)} } \nonumber \\
                & \mp \frac{9 \bar\gamma^2 \mathcal{A}^\prime(\Phi) + 4 \Pi \bar\gamma \mathcal{B}(\Phi)}{3 \left(4 \Pi \mathcal{B}(\Phi) + 3 \bar\gamma \mathcal{A}^\prime{\left(\Phi \right)}\right) } 
                \sqrt{\frac{3 (\Pi^{2} \mathcal{B}{\left(\Phi \right)} +3 \Pi \bar\gamma \mathcal{A}^\prime{\left(\Phi \right)} + 2 \kappa^{2} \rho + 2 \mathcal{V}{\left(\Phi \right)})}{2\mathcal{A}{\left(\Phi \right)}}} \,,
          \end{align}
   \end{subequations}
    where $\dot{\bar{\gamma}}$ in \eqref{eq: 1st order system set 3 Pi} should be substituted in from \eqref{eq: 1st order system set 3 gamma}. This can be compared to the systems \eqref{eq: 1st order system set 1} of set 1
     and \eqref{eq: 1st order system set 2} of set 2. There are four independent variables $\{\Phi, \dot{\Phi}, \rho, \gamma \}$, and for generic model functions it is not possible to reduce the system to a lower dimensional one.

     Concerning the scalar field dynamics, Eq.\ \eqref{eq: 1st order system set 3 Pi} is structurally rather similar to Eq.\ \eqref{eq: 1st order system set 2 Pi} of set 2 that we analyzed above. The only addition in \eqref{eq: 1st order system set 3 Pi} is another contribution to friction which depends on $\bar\gamma$. Therefore all the remarks about the possible instabilities also apply to the set 3 connection here, and we should be careful about that.

%%%%%%%%%%%%%%%%%%%%%%%%%%%%%%55
 \section{Limit of general relativity}
      \label{sec: Limit of GR}
Since the extra function in the FLRW symmetric teleparallel connections can not contribute to dark matter or dark energy, but rather induces a possibility for instabilities, it would make sense to check whether the usual $\Lambda$CDM background behavior is obtainable in the theory, and prepare ground to study the dynamics near such regime.
\subsection{Relaxation to general relativity} \label{subsec:Relaxation to general relativity}
 The well known Friedmann equations for spatially flat universe with a single barotropic fluid matter component in general relativity are
         \begin{subequations}
        \label{eq: GR Friedmann equations}
             \begin{align}
                 3 H^2 &= 8 \pi G_N \rho + \Lambda \,,\\
                 2 \dot{H} + 3 H^2 &= - 8\pi G_N \mathrm{w} \rho + \Lambda \,, \\
                \dot\rho + 3H (1+\mathrm{w}) \rho &=0 \,,
             \end{align}
    \end{subequations}
     where $G_N$ is the Newtonian gravitational constant and $\Lambda$ the cosmological constant. The scalar-tensor cosmological metric equations for different connections \eqref{eq: cosmology eq set 1 FR1}--\eqref{eq: cosmology eq set 1 FR2}, \eqref{eq: cosmology eq set 2 FR1}--\eqref{eq: cosmology eq set 2 FR2}, and \eqref{eq: cosmology eq set 3 FR1}--\eqref{eq: cosmology eq set 3 FR2} reduce to \eqref{eq: GR Friedmann equations} when the dynamics of the scalar field stops, i.e.\ at $\Phi_*$ which sustains $\dot{\Phi}=\ddot{\Phi}=0$. Then the value of the nonminimal coupling function sets the gravitational constant, $8 \pi G_N = \tfrac{\kappa^2}{\mathcal{A}(\Phi_*)}$, and the value of the potential plays the role of the cosmological constant, $\Lambda=\frac{\mathcal{V}(\Phi_*)}{\mathcal{A}(\Phi_*)}$. Note that the vanishing of $\dot\Phi$ also removes the contribution of the connection functions $\gamma$ and $\bar\gamma$ from the metric field equations.
        Whether the stabilization of the scalar field to a constant value is possible and at which value of $\Phi$ it does occur, depends on the model functions $\mathcal{A}(\Phi)$, $\mathcal{B}(\Phi)$ $\mathcal{V}(\Phi)$, and the respective connection and scalar field equations \eqref{eq: cosmology eq set 1 SE}, \eqref{eq: cosmology eq set 2 CE}--\eqref{eq: cosmology eq set 2 SE}, \eqref{eq: cosmology eq set 3 CE}--\eqref{eq: cosmology eq set 3 SE}.

        To shorten the expressions in the calculations that follow, it is helpful to remember that without the loss of generality we can reparametrize the scalar field, i.e.\ introduce new $\phi(\Phi)$ so that the general form of the the action \eqref{Action} does not change. Since by our starting assumptions the scalar field is only involved in mediating gravity and does not have other interactions with the matter fields, the precise value of the scalar field is unmeasurable and irrelevant. All that matters in the cosmological context is the coupling to the gravity term $Q$ in the action which sets the effective gravitational constant, and the kinetic and potential terms in the equations, which affect the dynamics. Thus for the sake of simplicity we can adopt a particular parametrization of the scalar field whereby  the kinetic coupling is canonical, $\mathcal{B}(\phi)\equiv 1$. We can also rewrite the nonminimal coupling to gravity as $\mathcal{A}(\phi) = 1 + f(\phi)$ which splits out a constant part, and denote the potential function in terms of the reparametrized field as $V(\phi)$.

        In terms of this parametrization the solutions of Eqs.\ \eqref{eq: GR Friedmann equations} for different matter types are easy to write out. In the case of nonrelativistic matter (pressureless dust, $\mathrm{w}=0$) the Hubble parameter and matter density evolve as
         \begin{align}
         \label{eq: dust dominant background}
             H_* &= \frac{2}{3(t-t_s)} \,, \qquad \rho_* = \frac{4(1+f_*)}{3 \kappa^2 (t-t_s)^2} \,, \qquad V_* = 0 \,,
     \end{align}
      in the case of relativistic matter (radiation, $\mathrm{w}=\tfrac{1}{3}$) they evolve as
     \begin{align}
        \label{eq: radiation dominant background}
             H_* &= \frac{1}{2(t-t_s)} \,, \qquad \rho_* = \frac{3(1+f_*)}{4 \kappa^2 (t-t_s)^2} \,, \qquad V_* = 0 \,,
     \end{align}
      and finally, in the case when the scalar potential plays the role of dark energy (cosmological constant) the evolution is
     \begin{align}
         \label{eq: dark energy dominant background}
               H_* &= \sqrt{\frac{V_*}{3(1+f_*)}} \,, \qquad \rho_* = 0 \,, \qquad V_* =\mathrm{const.}\neq 0 \,.
    \end{align}
      Here $H_*$ and $\rho_*$ can evolve in time, but $f_*=f(\phi_*)$ and $V_*=V(\phi_*)$ are constants evaluated at the point where the scalar field stops. The integration constant $t_s$ sets the moment of initial singularity, and we can fix $t_s=0$. The cosmological standard $\Lambda$CDM model involves all three types of matter, but since the densities of the matter types evolve at different rates, the universe goes through a sequence of radiation, dust matter, and dark energy domination eras where the other components can be neglected as subdominant.

\subsection{Expansion around the general relativity limit} \label{subsec:Expansion around the general relativity limit}
In a realistic scenario for the late universe the eras of radiation, dust, and dark energy domination follow each other as the respective energy densities become dominant in succession. Thus a basic stability test of a modified gravity model is whether the single fluid cosmological equations are stable against small perturbations, i.e.\ that there is no other phenomenon that could spoil the standard history of the universe. Moreover, the change of the effective gravitational constant has significant constraints since the the Big Bang nucleosynthesis time, hence the scalar must have resided at an almost constant value since early universe till today \cite{Will:2014kxa,Uzan:2010pm,Ooba:2016slp,Alvey:2019ctk,Ballardini:2021evv}.

     To see whether the limit of general relativity is dynamically stable, let us study the evolution of small perturbations near the value of $\phi_*$. Thus let us expand
     \begin{align}
        \label{eq: general perturbations}
              \phi(t)&=\phi_{*}+x(t) \,, \qquad H(t)=H_{*}(t)+h(t) \,, \qquad \gamma(t)=\gamma_*(t) + g(t) \,, \qquad \rho(t)=\rho_*(t) + r(t) \,, 
        %\label{eq.34}
     \end{align}
      where $x(t)$, $h(t)$, $g(t)$, and $r(t)$ are small perturbations which we assume to be of roughly the same order. The respective derivatives are 
      \begin{align}
          \label{eq: general perturbations of derivatives}
               \dot{\phi}(t)=\dot{x}(t) \,, \qquad \dot{H}(t)=\dot{H}_{*}(t)+\dot{h}(t) \,, \qquad\dot{\gamma}(t)=\dot{\gamma}_{*}(t)+\dot{g}(t) \,, \qquad\dot{\rho}(t)=\dot{\rho}_{*}(t)+\dot{r}(t) \,, %\label{eq.35}
      \end{align}
       also assumed to be of roughly the same order small.
      We can also expand the functions of the scalar field as
      \begin{subequations}
      \label{eq: general perturbations of functions}
           \begin{align}
               f(\phi(t)) &= f_{*}+f'_{*}x(t) + \tfrac{f''_* \left(x(t) \right)^2}{2}  \,, \qquad
               f'(\phi(t)) = f'_{*}+f''_{*}x(t) + \tfrac{f'''_* \left(x(t) \right)^2}{2}  \,, \qquad
               f''(\phi(t))=f''_{*}+f'''_{*}x(t) + \tfrac{f''''_* \left(x(t) \right)^2 }{2} \,,\\
                V(\phi(t))&=V_{*}+V'_{*}x(t) + \tfrac{V''_* \left(x(t) \right)^2}{2}  \,, \qquad
               V'(\phi(t))=V'_{*}+V''_{*}x(t) + \tfrac{V'''_* \left(x(t) \right)^2}{2}  \label{eq.36} 
           \end{align}
  \end{subequations}
  Substituting these definitions into the cosmological equations \eqref{eq.27},  \eqref{eq:cosmo equations set 2}, and \eqref{eq:cosmo equations set 3} yield expressions which can be solved order by order.  
     At the lowest (background) order, for all three sets the metric and matter equations reduce to \eqref{eq: GR Friedmann equations} as explained above. At the background order, for all the connections, the connection equation is identically satisfied, while the scalar field equation is slightly different. 
     
\section{Stability of the standard cosmological regimes} \label{sec:Stability of standard cosmological regimes}
 Now we are in a position to investigate the stability of the standard dust, radiation, and dark energy dominated regimes near the general relativity limit in the configurations with different symmetric teleparallel cosmological connections.
 
\subsection{Connection set 1} \label{subsec:Connection set 1}
 Applying the parametrization and expansion introduced in Sec.\ \ref{sec: Limit of GR} on the cosmolgical equations \eqref{eq.27} of connection set 1 \eqref{eq: connection set 1} yields up to the first order small quantities
      \begin{subequations}
            \begin{align}
       \label{eq: expanded cosmo set 1 FR1}
               6 (1+f_{*}) H_{*}^{2}(t) - 2 V_{*} - 2 \kappa^{2} \rho_{*}  (t) + \Big( 12 (1+f_{*})  H_{*}  (t) h  (t) + 6 f^{\prime}_*  H_{*}^{2}  (t) x  (t) - 2 V^{\prime}_* x(t) - 2 \kappa^{2} r(t) \Big) &= 0 \,, \\
      \label{eq: expanded cosmo set 1 FR2}
              4 (1+f_{*}) \dot H_{*}(t) + 6 (1+f_{*}) H_{*}^{2}(t) - 2 V_{*} + 2 \kappa^{2} \mathrm{w} \rho_{*}(t) + \Big( 4 f^{\prime}_* x(t) \dot H_{*}(t)  + 12 (1+ f_{*}) H_{*}(t) h(t) + 4 (1+f_{*}) \dot h(t) & \nonumber \\  + 6 f^{\prime}_* H_{*}^{2}(t) x(t) + 4 f^{\prime}_* H_{*}(t) \dot x(t) - 2 V^{\prime}_* x(t) + 2 \kappa^{2} \mathrm{w} r(t) \Big)  &= 0 \,, \\
      \label{eq: expanded cosmo set 1 SE}
              -6 f^{\prime}_* H_{*}^{2}(t) - 2 V^{\prime}_* 
              - \Big( 2 V^{\prime\prime}_* x(t) + 6 f^{\prime\prime}_* H_{*}^{2}(t) x(t) 
             + 12 f^{\prime}_* H_{*}(t) h(t) + 6 H_{*}(t) \dot x(t) + 2 \ddot x(t) \Big) &=0 \,, \\
     \label{eq: expanded cosmo set 1 ME}
             \dot \rho_{*}(t) + 3 (1+\mathrm{w}) H_{*}(t) \rho_{*}(t) + \Big( 3 \mathrm{w} H_{*}(t) r(t) + 3 \mathrm{w} \rho_{*}(t) h(t) + 3 H_{*}(t) r(t) + 3 \rho_{*}(t) h(t) + \dot r(t)  \Big) & = 0 \,.
         \end{align}
   \end{subequations}
   The characteristic behavior depends on the dominant matter type. The same equations but without matter perturbation $r(t)$ were analyzed in Ref.\ \cite{Jarv:2015odu}. Therefore below we add the matter perturbation and summarize the main results.
   
\subsubsection{Dust matter domination} \label{subsubsecDust matter domination}
   In a nonrelativistic ($\mathrm{w}=0$) matter dominated scenario, when the matter energy density surpasses the potential and we can ignore $V_*$ over $\rho(t)$, the leading order (background) expressions of Eqs.\ \eqref{eq: expanded cosmo set 1 FR1}, \eqref{eq: expanded cosmo set 1 FR2}, \eqref{eq: expanded cosmo set 1 ME} are solved by the standard dust dominated background \eqref{eq: dust dominant background}. The remaining leading order part of the scalar field equation \eqref{eq: expanded cosmo set 1 SE} then demands $f'_*=0$, $V'_*=0$. This means a necessary condition for the scalar field evolution to stop and the system to relax to the general relativity regime is that the functions of the gravitational coupling $f(\phi)$ and the potential $V(\phi)$ must have coincident critical points at the same value of $\phi_*$, otherwise the scalar field can not stabilize and the expansion around $\phi_*$ is meaningless. 

     Provided a scalar-tensor model has such model functions, we can substitute the expressions $H_*(t)$ and $\rho_*(t)$ \eqref{eq: dust dominant background} into the first order perturbed parts of Eqs.\ \eqref{eq: expanded cosmo set 1 FR1}--\eqref{eq: expanded cosmo set 1 ME}, and solving those find that the small perturbations of the Hubble parameter and matter density decay in time,
     \begin{align}
           h(t) \sim t^{-2} \,, \qquad r(t) \sim t^{-3} \,.
       \end{align} 

       Upon substituting the background value of $H_{*}(t)$ into the perturbed scalar field equation \eqref{eq: expanded cosmo set 1 SE}, we get a Klein-Gordon equation in curved spacetime,
       \begin{align} \label{eq.simplifiedm SE}
           \ddot{x}(t)+\frac{2}{t}\dot{x}(t)+\left(V_{*}^{\prime\prime}+\frac{4f_{*}^{\prime\prime}}{3t^{2}}\right)x(t) = 0 \,.
       \end{align}
       For large $t$ the only surviving mass term is $V_*''$ which predicts oscillatory solutions for the minimum of the potential, $V_*''>0$, and exponential growth for the maximum of the potential, $V_*''<0$. The term linear in $\dot{x}$ modifies this overall dynamics by adding extra frictional damping. In fact, the above equation \eqref{eq.simplifiedm SE} can be recognized as a Bessel equation and solved exactly in terms of the Bessel functions or modified Bessel functions. Ignoring the oscillating factors in the solutions, the leading time dependence of the scalar perturbation turns out to be \cite{Jarv:2015odu}
     \begin{align}
          x(t) \sim \left\lbrace \begin{array}{ll} t^{-1} \,, & V''_*>0 \\ t^{-\frac{1}{2}} \,, & V''_*=0 \,, \quad v_* \,\, \mathrm{imaginary}  \\ t^{-\frac{1}{2} \left(1-v_*\right)} \,, & V''_*=0 \,, \quad v_* \,\, \mathrm{real} \\ t^{-1} e^{{\sqrt{-V''_*}}t} \,, & V''_*<0 \,. \end{array}\right.  \,, \qquad \quad v_*= \sqrt{1-\tfrac{16 f''_*}{3}}
      \end{align}
  Thus in summary, for the perturbations to converge and the general relativistic dust dominated regime to be stable, the model functions must have the value $\phi_*$ which corresponds to either a simultaneous local minimum of the potential and a critical point of the gravitational coupling function ($V'_*=0, V''_*>0, f'_*=0$), or to a local minimum of the gravitational coupling function and inflection point of the potential ($f'_*=0, f''_*>0, V'_*=0, V''_*=0$). 
\subsubsection{Radiation domination} \label{subsubsec:Radiation domination}
    In a relativistic $\left(\mathrm{w}=\tfrac{1}{3}\right)$ matter dominated case, when we can ignore $V_*$ in the presence of $\rho(t)$, the leading order expressions of Eqs.\ \eqref{eq: expanded cosmo set 1 FR1}, \eqref{eq: expanded cosmo set 1 FR2}, \eqref{eq: expanded cosmo set 1 ME} are solved by the standard radiation dominated background \eqref{eq: radiation dominant background}. The remaining leading order part of the scalar field equation \eqref{eq: expanded cosmo set 1 SE} then again demands $f'_*=0$, $V'_*=0$.
        Provided the scalar-tensor model has such suitable model functions, we can substitute the expressions $H_*(t)$ and $\rho_*(t)$ from \eqref{eq: radiation dominant background} into the first order perturbed parts of Eqs.\ \eqref{eq: expanded cosmo set 1 FR1}--\eqref{eq: expanded cosmo set 1 ME} and find that the small perturbations of the Hubble parameter and matter density decay in time,
         \begin{align}
         h(t) \sim t^{-2} \,, \qquad r(t) \sim t^{-3} \,.
      \end{align} 
      After substituting the background value $H_{*}(t)$  into equation \eqref{eq: expanded cosmo set 1 SE}, we again get a Klein-Gordon type equation,
      \begin{align} 
      \label{eq.simplifiedr SE}
           \ddot{x}(t)+\frac{3}{2t}\dot{x}(t)+\left(V_{*}^{\prime\prime}+\frac{3f_{*}^{\prime\prime}}{4t^{2}}\right) x(t) = 0 \,.
       \end{align}
    % The first order perturbed scalar field equation takes again 
    The overall stability is principally determined by the sign of $V_*''$, i.e.\ whether we are at the maximum or minimum of the potential, while the $\dot{x}$ term adds an extra friction effect. The full solutions to the equation \eqref{eq.simplifiedr SE} can be again found in the form of a the Bessel functions,
     whereby ignoring the oscillating factors the leading time dependence of the scalar perturbation turns out to be \cite{Jarv:2015odu}
    \begin{align}
           x(t) \sim \left\lbrace \begin{array}{ll} t^{-\frac{3}{4}} \,, & V''_*>0 \\ t^{-\frac{1}{4}} \,, & V''_*=0 \,, \quad v_* \,\, \mathrm{imaginary} \\ t^{-\frac{1}{4}\left(1-v_*\right)} \,, & V''_*=0 \,, \quad v_* \,\, \mathrm{real} \\ t^{-\frac{3}{4}} e^{{\sqrt{-V''_*}}t} \,, & V''_*<0 \,. \end{array}\right. \,, \qquad \quad  v_*=\sqrt{1-12f''_*} \,.
      \end{align}
       Thus in summary, just like in the dust matter case, for the perturbations to converge and the general relativistic radiation dominated regime to be stable, the model functions must have the value $\phi_*$ which corresponds to either a simultaneous local minimum of the potential and a critical point of the gravitational coupling function ($V'_*=0, V''_*>0, f'_*=0$), or to a local minimum of the gravitational coupling function and inflection point of the potential ($f'_*=0, f''_*>0, V'_*=0, V''_*=0$).
       
 \subsubsection{Potential domination} \label{subsubsec:Potential domination}
      In the era when the scalar potential dominates over the matter energy density, and we can drop $\rho(t)$ in comparison with $V({\phi_*})$ (but keep dust matter perturbations $r(t)$ with $\mathrm{w}=0$) the leading order expressions of Eqs.\ \eqref{eq: expanded cosmo set 1 FR1}, \eqref{eq: expanded cosmo set 1 FR2}, \eqref{eq: expanded cosmo set 1 ME} are solved by the standard dark energy (cosmological constant) dominated background \eqref{eq: dark energy dominant background}. The remaining leading order part of the scalar field equation \eqref{eq: expanded cosmo set 1 SE} can then be satisfied in two ways. 

     First, the background scalar field equation can be solved by $f'_*=0$, $V'_*=0$. Then the first order perturbed metric and matter equations give 
     \begin{align}
          h(t) \sim e^{-3H_* t} \,, \qquad  r(t) \sim e^{-3H_* t} \,.
     \end{align}
     The first order perturbed scalar field equation 
     \begin{align} 
      \label{eq.simplifiedH SE}
           \ddot{x}(t)+3H_*\dot{x}(t)+\left(V_{*}^{\prime\prime}+3 H_*^2 f_{*}^{\prime\prime}\right) x(t) = 0 \,
       \end{align}
     is again a Klein-Gordon type with a friction term. The overall stability, i.e.\ exponentially damped oscillations or exponential growth depends on the sign of the third term (effective mass squared). Dropping the factor of oscillations in the solutions, the time dependence of the scalar field perturbation can be expressed as \cite{Jarv:2015odu}
      \begin{align}
          x(t) \sim \left\lbrace \begin{array}{ll} e^{-\frac{3 H_* t}{2}} \,, & v_* \,\, \mathrm{imaginary} \\e^{-\frac{3 H_* t}{2}(1-v_*)} \,, & v_* \,\, \mathrm{real} \, \end{array}\right. \,, \qquad 
          v_*=\sqrt{1- \tfrac{4}{9 H_*^2} \left(V_*'' + 3 H_*^2 f_*'' \right)}
          %v_*=\sqrt{1-\tfrac{4 f''_*}{3} - \tfrac{4(1+f_*)V''_*}{3V_*}}
     \end{align}
     Thus to realize a stable dark energy era in this scenario, the model functions must have the value $\phi_*$ which corresponds to either a local minimum of the potential and simultaneously to a local minimum of the gravitational coupling function ($V'_*=0, V''_*>0, f'_*=0, f''_*>0$), or at least to a simultaneous critical point of the potential and the gravitational coupling ($f'_*=0, V'_*=0$) with an additional condition $V_* f''_* + (1+f_*) V''_*>0$.

  The background scalar field equation \eqref{eq: expanded cosmo set 1 SE} can be also satisfied by a ``balanced'' configuration where the scalar field resides not at a critical point of the potential or of the gravitational coupling function, but at a value which satisfies $V_*f'_* = - V'_*(1+f_*)$ whereby Eq.\ \eqref{eq: dark energy dominant background} implies $H_*=\sqrt{-\tfrac{V_*'}{3 f_*'}}$. In this case the perturbed part of the matter continuity equation gives \eqref{eq: expanded cosmo set 1 ME}
   \begin{align}
        r(t) \sim e^{-3 H_* t} \,, 
    \end{align} 
   while the remaining perturbed equations show that taking the leading orders the Hubble and scalar field perturbations are proportional to each other, $h(t) \sim x(t)$. The latter evolves according to a Klein-Gordon type equation
   \begin{align} 
      \label{eq.simplifiedHb SE}
           \ddot{x}(t)+3H_*\dot{x}(t)+\left(V_{*}^{\prime\prime}+3 H_*^2 f_{*}^{\prime\prime} + 2 \frac{(V_*')^2}{V_*}\right) x(t) = 0 \,. 
   \end{align}
   The behavior of the solutions of this equation again depends on the sign of the last term, and can be summarized as \cite{Jarv:2015odu}
   \begin{align}
       h(t) \sim x(t) \sim \left\lbrace \begin{array}{ll} e^{-\frac{3 H_* t}{2}} \,, & v_* \,\, \mathrm{imaginary} \\e^{-\frac{3 H_* t}{2}(1-v_*)} \,, & v_* \,\, \mathrm{real} \, \end{array}\right. \,, \qquad 
       v_*=\sqrt{1- \tfrac{4}{9 H_*^2} \left( V''_* + 3 H_*^2 f_*'' + \tfrac{2 (V_*')^2}{V_*} \right)} \,.
%       v_*=\sqrt{1-\tfrac{4 f''_*}{3} - \tfrac{4(1+f_*)V''_*}{3V_*} - \tfrac{8f'_*V'_*}{V_*}}
   \end{align}
    The perturbations die off in time and this configuration is stable if $V_* f''_* + (1+f_*) V'' + 2 f'_* V'_* >0$. However, in constructing a realistic cosmic history, we might prefer this configuration to be unstable like a saddle point, and play a role as a launching point for an inflationary trajectory \cite{Jarv:2021ehj}. If it is unstable, it can trigger an early epoch of inflationary expansion making the scalar field to roll to a value where $f'_*=0$, $V'_*=0$. If the latter point is stable, inflation would give way to radiation, dust matter and dark energy eras whereas the scalar field just relaxes in damped oscillations around that second point.

   \subsection{Connection set 2} \label{subsec:Connection set 2}
     Applying the parametrization and expansion introduced in Sec.\ \ref{sec: Limit of GR} on the cosmolgical equations \eqref{eq:cosmo equations set 2} of connection set 2 yields
   \begin{subequations}
       \label{eq: expanded cosmo set 2}
            \begin{align}
      \label{eq: expanded cosmo set 2 FR1}
                   6 (1+f_{*}) H_{*}^{2}(t) - 2 V_{*} - 2 \kappa^{2} \rho_{*}  (t) + \Big( 12 (1+f_{*})  H_{*}  (t) h  (t) + 6 f^{\prime}_*  H_{*}^{2}  (t) x  (t) + 3 f^{\prime}_* \gamma_{*}(t)  \dot x(t) - 2 V^{\prime}_* x(t) - 2 \kappa^{2} r(t) \Big) &= 0 \,, \\
      \label{eq: expanded cosmo set 2 FR2}
                   4 (1+f_{*}) \dot H_{*}(t) + 6 (1+f_{*}) H_{*}^{2}(t) - 2 V_{*} + 2 \kappa^{2} \mathrm{w} \rho_{*}(t) + \Big( 4 f^{\prime}_* x(t) \dot H_{*}(t)  + 12 (1+ f_{*}) H_{*}(t) h(t) + 4 (1+f_{*}) \dot h(t) & \nonumber \\  + 6 f^{\prime}_* H_{*}^{2}(t) x(t) + 4 f^{\prime}_* H_{*}(t) \dot x(t) - 3 f^{\prime}_* \gamma_{*}(t) \dot x(t) - 2 V^{\prime}_* x(t) + 2 \kappa^{2} \mathrm{w} r(t) \Big)  &= 0 \,, \\
       \label{eq: expanded cosmo set 2 CE}
                  \Big( 9 f^{\prime}_* H_{*}(t) \gamma_{*}(t) \dot x(t) + 3 f^{\prime}_* \gamma_{*}(t) \ddot x(t) \Big)
                  + \Big( 9 f^{\prime\prime}_* H_{*}(t) \gamma_{*}(t) x(t) \dot x(t) + 3 f^{\prime\prime}_* \gamma_{*}(t) x(t) \ddot x(t) + 3 f^{\prime\prime}_* \gamma_{*}(t) \left(\dot x(t)\right)^{2} & \nonumber \\ 
                   + 9 f^{\prime}_* H_{*}(t) g(t) \dot x(t) + 9 f^{\prime}_* \gamma_{*}(t) h(t) \dot x(t) + 3 f^{\prime}_* g(t) \ddot x(t) \Big) &=0 \,, \\
       \label{eq: expanded cosmo set 2 SE}
                 3 f^{\prime}_* \left( \dot \gamma_{*}(t) - 2 H_{*}^{2}(t) + 3 H_{*}(t) \gamma_{*}(t) \right) - 2 V^{\prime}_* 
                - \Big( 2 V^{\prime\prime}_* x(t) + 6 f^{\prime\prime}_* H_{*}^{2}(t) x(t) - 9 f^{\prime\prime}_* H_{*}(t) \gamma_{*}(t) x(t) - 3 f^{\prime\prime}_* x(t) \dot \gamma_{*}(t)  & \nonumber \\ 
                 + 12 f^{\prime}_* H_{*}(t) h(t) - 9 f^{\prime}_* \gamma_{*}(t) h(t) - 9 f^{\prime}_* H_{*}(t) g(t) - 3 f^{\prime}_* \dot g(t) + 6 H_{*}(t) \dot x(t) + 2 \ddot x(t) \Big) &=0 \,, \\
       \label{eq: expanded cosmo set 2 ME}
                \dot \rho_{*}(t) + 3 (1+\mathrm{w}) H_{*}(t) \rho_{*}(t) + \Big( 3 \mathrm{w} H_{*}(t) r(t) + 3 \mathrm{w} \rho_{*}(t) h(t) + 3 H_{*}(t) r(t) + 3 \rho_{*}(t) h(t) + \dot r(t)
               \Big) & = 0 \,.
           \end{align}
  \end{subequations}
   We will consider the stability in the case of different matter types separately, going through the dust matter calculations in detail, and presenting the main results in the other cases.
   
\subsubsection{Dust matter domination} 
    \label{subsubsec:Dust matter domination conn 2}
       The regime of dust matter domination means we take $\mathrm{w}=0$ and neglect $V_*$ in comparison with $\rho(t)$. At the background level (without perturbations) the metric and matter equations \eqref{eq: expanded cosmo set 2 FR1}, \eqref{eq: expanded cosmo set 2 FR2}, \eqref{eq: expanded cosmo set 2 ME} coincide with the GR cosmological equations \eqref{eq: GR Friedmann equations} and are solved by the standard background evolution \eqref{eq: dust dominant background}. The background part of the connection equation \eqref{eq: expanded cosmo set 2 CE} is identically zero, while in solving the scalar field equation \eqref{eq: expanded cosmo set 2 SE} at the background level,
       \begin{align}
           3 f'_* \left(\dot{\gamma}_*(t) + \frac{ 2 \gamma_*(t) }{t} -\frac{8}{9 t^2} \right) - 2 V_*' &=0 \label{eq: dust bg gamma eq} \,,
       \end{align}
     there are three options. Either $f_*' \neq 0$, $V_*' \neq 0$ and the evolution of $\gamma_*(t)$ is given by
       \begin{align}
         \label{eq: dust bg gamma sol}
               \gamma_*(t) &= \frac{2 V^{\prime}_* t }{9 f^{\prime}_*} + \frac{8}{9t} + \frac{c_1}{t^2}
     \end{align}
     which leaves the value of $\phi_*$ arbitrary; or $f_*' \neq 0$, $V_*' = 0$ in \eqref{eq: dust bg gamma eq} and \eqref{eq: dust bg gamma sol}; or the model functions $f(\phi)$ and $V(\phi)$ have both an extremum at the same value of $\phi$ and the respective derivatives vanish,
     \begin{align}
       \label{eq: dust bg f V sol}
               f_*' &= 0 \,, \qquad V_*' = 0 \,,
    \end{align}
    which fixes $\phi_*$ but leaves $\gamma_*(t)$ undetermined. 

    In the first case, substituting the background values \eqref{eq: dust dominant background} and \eqref{eq: dust bg gamma sol} into the first order perturbed equations, i.e.\ keeping only terms that are linear in $x(t)$, $h(t)$, $g(t)$, and $r(t)$ in \eqref{eq: expanded cosmo set 2} gives   
  \begin{subequations}
       \begin{align}
               \frac{4 \left(1+f_{*}\right) h{\left(t \right)}}{t} - \frac{\left(3 V^{\prime}_* t^{2} - 4 f^{\prime}_*\right) x{\left(t \right)}}{3 t^{2}} + \frac{\left(2 V^{\prime}_* t^{3} + f^{\prime}_* \left(9 c_{1} + 8 t\right)\right) \dot x{\left(t \right)}}{6 t^{2}} &= \kappa^{2} r{\left(t \right)} \,, \label{eq: dust 1p fr eq} \\
               - 4 \left(1+f_{*}\right) \dot h{\left(t \right)} - \frac{8 \left(1+f_{*}\right) h{\left(t \right)}}{t} + \frac{\left(2 V^{\prime}_* t^{3} + 9 c_{1} f^{\prime}_*\right) \dot x{\left(t \right)}}{3 t^{2}} + 2 V^{\prime}_* x{\left(t \right)} &=0 \,, \label{eq: dust 1p h eq} \\
                \frac{\left(2 V^{\prime}_* t^{3} + 9 c_{1} f^{\prime}_* + 8 f^{\prime}_* t\right)}{3 t^{3}} \left(t \ddot x{\left(t \right)} + 2 \dot x{\left(t \right)}\right) &=0 \,. \label{eq: dust 1p connection eq} \\
               f^{\prime}_* \left(3 \dot g{\left(t \right)} + \frac{6 g{\left(t \right)}}{t}\right) + \frac{\left(2 V^{\prime}_* t^{3} + 9 c_{1} f^{\prime}_*\right) h{\left(t \right)}}{t^{2}} - \frac{2 \left(V^{\prime\prime}_* f^{\prime}_* - V^{\prime}_* f^{\prime\prime}_*\right) x{\left(t \right)}}{f^{\prime}_*} - 2 \ddot x{\left(t \right)} - \frac{4}{t} \dot x(t)  &=0 \,, \label{eq: dust 1p g eq} \\
          \label{eq: dust 1p r eq}
                 \dot r(t) + \frac{2 r(t)}{t} + \frac{4(1+f_*)h(t)}{\kappa^2 t^2} &=0 \,.
      \end{align}
  \end{subequations}
  Here we can first integrate Eq.\ \eqref{eq: dust 1p connection eq}, to get 
         \begin{align}
            \label{eq: dust 1p x sol}
               x(t) &= \frac{c_2}{t} + c_3 \,.
         \end{align}
         The integration constant $c_3=0$, since we have defined that in the end the scalar field stops at $\phi_*$, not at $\phi_*+c_3$.
    Thus as the time passes on, the perturbation in the scalar field diminishes, which ensures that the $\gamma(t) \dot \phi(t)$ term in the Friedmann equation  \eqref{eq: expanded cosmo set 2 FR1} does not grow to spoil the dust domination regime, despite $\gamma(t)$ \eqref{eq: dust bg gamma sol} increasing in time. Substituting \eqref{eq: dust 1p x sol} into \eqref{eq: dust 1p h eq} we get an equation for $h(t)$, which is solved by
     \begin{align}
             \label{eq: dust 1p h sol}
                  h(t) &= \frac{ c_{2} V^{\prime}_*}{6 \left(1+f_{*}\right)} + \frac{c_{4}}{ t^{2}} + \frac{3 c_{1} c_{2} f^{\prime}_*}{4 \left(1+f_{*}\right) t^{3}} \,.
        \end{align}
          Then we can substitute \eqref{eq: dust 1p x sol} and \eqref{eq: dust 1p h sol} into \eqref{eq: dust 1p g eq} and solve to find
   \begin{align}
                g(t) &= - \frac{c_{2} \left(V^{\prime}_*\right)^{2} t^{2}}{36 f^{\prime}_* \left(1+f_{*}\right)} 
                + \frac{c_{2} V^{\prime\prime}_* f^{\prime}_* - c_{2} V^{\prime}_* f^{\prime\prime}_* - c_{4} V^{\prime}_* f^{\prime}_*}{3\left(f^{\prime}_*\right)^{2}}
                - \frac{c_{1} c_{2} V^{\prime}_*}{\left(1+f_{*}\right) t} 
                + \frac{c_{5}}{t^{2}} 
                + \frac{3c_{1} c_{4}}{ t^{3}} 
                + \frac{9 c_{1}^{2} c_{2} f^{\prime}_*}{8 \left(1+f_{*}\right) t^{4}} \,.
        \end{align}
        Finally the solutions \eqref{eq: dust 1p x sol} and \eqref{eq: dust 1p h sol} can be substituted into \eqref{eq: dust 1p r eq} from where it is possible to algebraically express
    \begin{align}
                r(t) &=- \frac{2 c_{2} V^{\prime}_*}{3 \kappa^{2} t} + \frac{4c_{4} \left(1+f_{*}\right)}{3 \kappa^{2} t^{3}} + \frac{3 c_{1} c_{2} f^{\prime}_*}{2 \kappa^{2} t^{4}}\,.
    \end{align}
           The latter expression along with with \eqref{eq: dust 1p h sol} also solves \eqref{eq: dust 1p r eq}.
   Thus we see, that if we perturb around an arbitrary scalar field value, where neither $V'_*\neq 0$ nor $f'_*\neq 0$, then to the leading order in time the quantities evolve as
     \begin{align}
              \gamma_*(t) & \sim t \,, \qquad x(t) \sim t^{-1} \,, \qquad h(t) \sim t^0 \,, \qquad g(t) \sim t^2 \,, \qquad r(t) \sim t^{-1} \,,
     \end{align}
           and the configuration can not be considered stable. Although the evolution of $\phi$ slows to a stop, and $\rho$ converges to its general relativity regime, and even the effects of the growing connection function are surpressed by decreasing $\dot \phi$, the connection function perturbations $g(t)$ grow bigger in time and eventually spoil the approximation which assumes the perturbations to be small. 

If the perturbation takes place around $f'_*\neq 0$, $V'_*=0$ mentioned as the second case above, then for $V''_* \neq 0$
         \begin{align}
                 \gamma_*(t) & \sim t^{-1} \,, \qquad x(t) \sim t^{-1} \,, \qquad h(t) \sim t^{-2} \,, \qquad g(t) \sim t^0 \,, \qquad r(t) \sim t^{-3} \,.
          \end{align}
Although the connection function perturbations do not decrease, they do not increase either, and the configutation may be considered marginally stable. If the second derivatives of the potential are also zero (like for a cosmological constant), then 
         \begin{align}
               \gamma_*(t) & \sim t^{-1} \,, \qquad x(t) \sim t^{-1} \,, \qquad h(t) \sim t^{-2} \,, \qquad g(t) \sim t^{-2} \,, \qquad r(t) \sim t^{-3} \,.
         \end{align}
         and a stable regime is possible.

\begin{figure}
    \centering
    \subfigure[]{
    \includegraphics[width=0.30\textwidth]{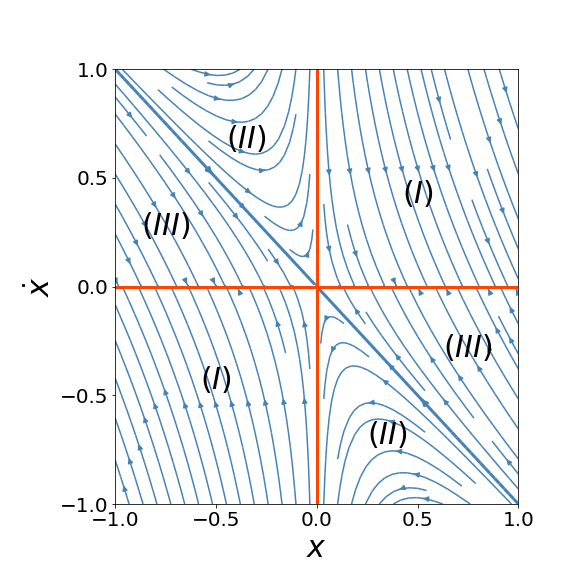}
    \label{fig: phase portrait set2 dust}}
    \subfigure[]{
    \includegraphics[width=0.30\textwidth]{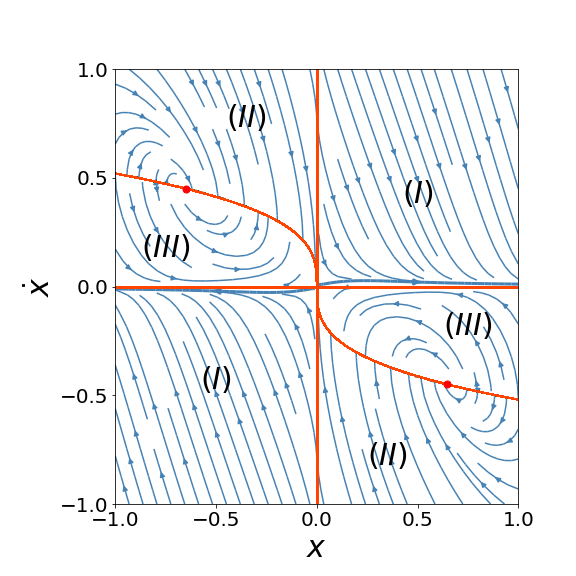}
    \label{fig: phase portrait set3b dust}}
    \subfigure[]{
    \includegraphics[width=0.30\textwidth]{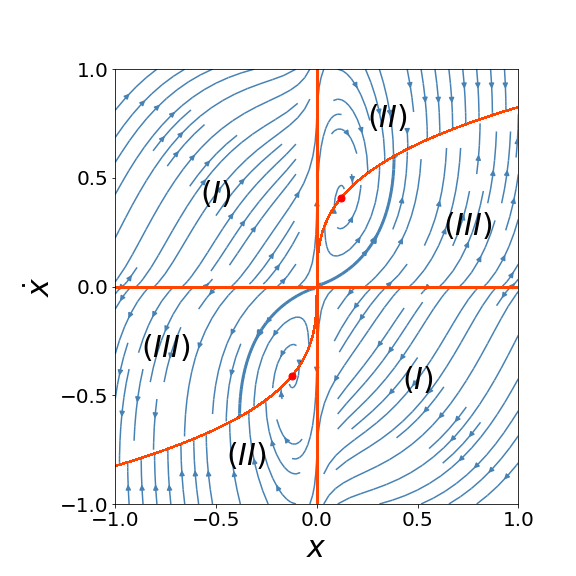}
    \label{fig: phase portrait set3c dust}}
    \caption{A sketch of the phase space where singular solutions in occur. The portrait of (a) set 2 dust dominated Eq.\ \eqref{eq: dust x_s(t) set 2 equation} at $t=1$, populated by the solutions \eqref{eq: dust x_s(t) set 2}; (b) set 3 dust dominated Eq.\ \eqref{eq: dust x_s(t) set 3 equation} for $V_*''=1$, $f_*''=1$, $c_9=0.5$ at $t=1$; (c) set 3 dust dominated Eq.\ \eqref{eq: dust x_s(t) set 3 equation} for $V_*''=1$, $f_*''=1$, $c_9=-1$ at $t=1$.}
    \label{fig: phase portraits}
\end{figure}

When we want to consider perturbations around a simultaneous critical point of the potential and gravitational coupling functions, Eq.\ \eqref{eq: dust bg f V sol}, the scalar field equation is identically solved at the background level and $\gamma_*(t)$ remains undetermined. The lowest order perturbed equations \eqref{eq: expanded cosmo set 2} are now
   \begin{subequations}
   \label{eq: dust 1pp}
         \begin{align}
               \frac{4 \left(1+f_{*}\right) h{\left(t \right)}}{t} &= \kappa^{2} r{\left(t \right)} \,, \label{eq: dust 1pp r eq}\\
               - 2 \left(1+f_{*}\right) \left(\dot{h}{\left(t \right)} + \frac{2 h{\left(t \right)}}{t}\right) &=0 \,, \label{eq: dust 1pp h eq} \\
          \label{eq: dust x_s(t) set 2 equation}
               x(t) \ddot{x}(t) + \dot{x}(t)^2 + \frac{2 x(t) \dot{x }(t)}{t} &=0 \,, \\
               \label{eq: dust1 SE eq}
               \left(9 f^{\prime\prime}_* t^{2} \dot{\gamma}_{*}{\left(t \right)} - 6 V^{\prime\prime}_* t^{2} + 18 f^{\prime\prime}_* t \gamma_{*}{\left(t \right)} - 8 f^{\prime\prime}_*\right) \frac{ x{\left(t \right)}}{3 t^{2}} - 2 \ddot x(t) - \frac{4}{t} \dot x(t) &=0 \,, \\
          \label{eq: dust 1pp x eq} 
               \dot r(t) + \frac{2 r(t)}{t} + \frac{4(1+f_*)h(t)}{\kappa^2 t^2} &=0 \,.
        \end{align}
    \end{subequations}
      Note that at the first (linear) perturbation level the connection equation is identically satisfied and \eqref{eq: dust x_s(t) set 2 equation} represents the second (nonlinear) perturbation level as the leading nonzero order. The equation for the Hubble perturbation \eqref{eq: dust 1pp h eq} can be integrated to
         \begin{align}
               h(t) &= \frac{c_2}{t^2} 
          \end{align}
           and Eq.\ \eqref{eq: dust 1pp r eq} then gives
          \begin{align}
               r(t) &= \frac{4 c_2 (1+f_*)}{\kappa^2 t^3} \,.
          \end{align}
We can not tackle the remaining first order perturbed scalar field equation \lptext{\eqref{eq: dust1 SE eq}}
% \eqref{eq: dust 1pp x eq} 
directly, since it contains two unknown functions $\gamma_*(t)$ and $x(t)$. However, it is still possible to proceed by taking the next order perturbation of the connection equation which contains second order small quantities \eqref{eq: dust x_s(t) set 2 equation}. Interestingly, the structure of that equation parallels the perturbed scalar field equation near the general relativity limit of the usual curvature based scalar-tensor cosmology with dust matter or potential \cite{Jarv:2011sm, Jarv:2010zc, Jarv:2010xm}. However, the signs are different, and the solutions behave differently here. The equation \eqref{eq: dust x_s(t) set 2 equation} can be solved easily by 
         \begin{align}
              \label{eq: dust x_s(t) set 2}
                     x(t) &= \pm \sqrt{\frac{c_6}{t}+c_7} \,.
         \end{align}
Despite the innocent looks, this expression actually harbors a singularity, related to the feature briefly discussed already in the end of Sec.\ \ref{subsec: cosmo equations of set 2}. We can compute from \eqref{eq: dust x_s(t) set 2} 
       \begin{align}
             \dot x(t) &= \mp \frac{c_6}{2 t^2 \sqrt{\frac{c_6}{t}+c_7}}
       \end{align}
        and express the integration constants in terms of initial conditions $x_0$, $\dot x_0$ at $t_0$ as
       \begin{align}
       \label{eq: set 2 particular solution}
             c_6 &= - 2 \dot x_0 \, x_0 \, t_0^2   \,, \qquad
             c_7 = x_0^2 + 2 \dot x_0 \, x_0 \, t_0  \,,
       \end{align}
        Both integration constants are real. At a finite time $t_*=-\frac{c_6}{c_7}$ the scalar perturbation $x(t)$ goes to zero, but the speed $\dot x(t)$ becomes singular and the approximation of perturbations being small breaks down. The singularity occurs in the physical time $t>0$ if $c_6$ and $c_7$ have opposite signs. We can write 
       \begin{align}
           \frac{t_0}{t_*} &= -\frac{c_7 t_0}{c_6} = 1+ \frac{x_0}{2\dot{x}_0t_0}
        \end{align}
        and analyze the situation in conjunction with the phase portrait on Fig.\ \ref{fig: phase portrait set2 dust} as follows. First, in regions $(I)$, where $x_0$ and $\dot x_0$ are of the same sign, then $t_*<t_0$ and the singularity happens in the past of the solution, i.e.\ the solution which is specified by some initial conditions at the moment $t_0$ has only emerged from singularity at a finite time ago. These solutions evolve away from $\phi_*$ and could be classified as unstable. Asymptotically they would reach $|x(t\to \infty)| \to \sqrt{c_7}$, but before that the current approximation breaks down, and they should fall under the purview of the more general case that follows from \eqref{eq: dust bg gamma sol} which was considered before. Second, in regions $(II)$, where $x_0$ and $\dot x_0$ are of opposite signs and $|\dot x_0| > \tfrac{|x_0|}{2 t_0}$, then $t_* > t_0$ and the singularity happens in the future of the solution. Definitely the approximation scheme breaks down as $\dot x$ diverges, but quite likely the full system hits a singularity. Thirdly, in regions $(III)$, where $x_0$ and $\dot x_0$ are of opposite signs but $|\dot x_0| < \tfrac{|x_0|}{2 t_0}$, the integration constants are of the same sign, and the solutions have a tendency to slow down and arrive at $|x(t\to \infty)| \to \sqrt{c_7}$. However, in the strict sense they would again belong to the more general case considered above, since the point of eventual stability for $\phi$ does not satisfy the condition \eqref{eq: dust bg f V sol}. Finally, there is a particular set of solutions specified by $x_0 = - 2 \dot x_0 t_0$ so that $c_7=0$. Running between regions $(II)$ and $(III)$ they come from $x>0$ and $x<0$. Only this type of solutions manage to asymptotically reach $|x(t\to \infty)| \to 0$ while avoiding the singularity as well as the fate of stopping before that value. To complete the analysis of this particular case we can substitute \eqref{eq: dust x_s(t) set 2} with $c_7=0$ into \eqref{eq: dust 1pp x eq} and find the background evolution of the connection function to be
        \begin{align}
              \gamma_*(t) &= \frac{2 V^{\prime\prime}_* t}{9 f^{\prime\prime}_*} + \frac{8}{9 t} - \frac{1}{6 f^{\prime\prime}_* t}  + \frac{c_{8}}{t^{2}} \,.
         \end{align}
           For nonzero $V_*''$ the connection function diverges linearly in time, but the $\gamma_* \dot \phi$ term in the Friedmann equation is still suppressed.  Although the Hubble, matter, and scalar field perturbations diminish in time, we can not safely conclude this particular type of solutions is convergent, since the connection perturbation $g(t)$ remains undetermined.
           
At this point an astute reader may raise a concern whether it was consistent to consider quantities quadratic in the perturbations only in the perturbed connection equation \eqref{eq: dust x_s(t) set 2 equation}, but not in the others of \eqref{eq: dust 1pp}. The reason is that for each equation we are interested only in the leading dominant behavior that is relevant for the stability of the system, i.e.\ whether the solutions converge to or diverge from the general relativity limit. Although in the other equations the subdominant quadratic and higher terms are also present, under the assumption of smallness they have less influence and do not decide the issue of stability. If we wanted to know the higher corrections to the time dependence of the solutions, then we would have needed to include also higher order small perturbations in the expansions \eqref{eq: general perturbations}. However, most likely the system of equations would then become even more complicated and harder to solve.

In any case, for large enough initial velocities, the system meets a singularity in finite time. It is remarkable, that the limit $V'_*=0$, $f'_*=0$ which was giving a stable standard history in the case of connection set 1 is unstable for a large range of initial conditions in the dust dominant case of set 2. The strange feature that the background equations fail to determine the connection function $\gamma_*(t)$ in this limit while the derivative of the scalar field perturbation becomes singular may indicate that the scalar-nonmetricity theory is problematic and does not reduce to general relativity in a smooth manner here. Alternatively, we may interpret this feature as an indication that the connection set 2 is unphysical and should be discarded in favor of the set 1 where the GR limit is smooth. In view of the remark in the end of Sec.\ \ref{subsec: cosmo equations of set 2} a more thorough investigation of the instability of the full equations might shed light on this issue, but it remains beyond the scope of the present work.

\subsubsection{Radiation domination} \label{subsubsec:Radiation domination for conn 2}

The regime of radiation domination allows us to neglect $V_*$ in comparison with $\rho(t)$ which is characterized by $\mathrm{w}=\tfrac{1}{3}$. The calculations can be performed analogously to the previous subsection, and we will just summarize the key results.
At the background level the metric and matter equations \eqref{eq: expanded cosmo set 2 FR1}, \eqref{eq: expanded cosmo set 2 FR2}, \eqref{eq: expanded cosmo set 2 ME} coincide with the GR cosmological equations \eqref{eq: GR Friedmann equations} and are solved by the standard background evolution \eqref{eq: radiation dominant background}. The background part of the connection equation \eqref{eq: expanded cosmo set 2 CE} is identically zero, while solving the scalar field equation \eqref{eq: expanded cosmo set 2 SE} at the background level,
        \begin{align}
              3 f'_* \left(\dot{\gamma}_*(t) + \frac{3 \gamma_*(t) }{2t} -\frac{1}{2 t^2} \right) - 2 V_*' &=0 \label{eq: radiation bg gamma eq} \,,
       \end{align}
        gives three options. 

  First, for an arbitrary $\phi_*$ whereby $f_*' \neq 0$, $V_*' \neq 0$, the evolution of $\gamma_*(t)$ in \eqref{eq: radiation bg gamma eq} is given by
   \begin{align}
          \label{eq: radiation bg gamma sol}
               \gamma_*(t) &= \frac{4 V^{\prime}_* t }{15 f^{\prime}_*} + \frac{1}{t} + \frac{c_1}{t^\frac{3}{2}}\,.
       \end{align}
       Substituting the background values \eqref{eq: radiation dominant background} and \eqref{eq: radiation bg gamma sol} into the first order perturbed parts of \eqref{eq: expanded cosmo set 2} gives a system of equations to determine the behavior of perturbations, which in the leading order turn out to be
   \begin{align}
              x(t) \sim t^{-\frac{1}{2}} \,, \qquad h(t) \sim t^\frac{1}{2} \,, \qquad  g(t) \sim t^\frac{5}{2} \,, \qquad r(t) \sim t^{-\frac{1}{2}}  \,.
       \end{align}
        Even the perturbations of the Hubble parameter grow, and the configuration is unstable.
        
  Second, if $\phi_*$ is fixed by $V_*' = 0$, but $f_*' \neq 0$, then the scalar field equation \eqref{eq: radiation bg gamma eq} is solved by the subleading terms in \eqref{eq: radiation bg gamma sol}. Then the first order perturbations in the leading order behave as 
          \begin{align}
             \gamma_{*}(t) \sim t^{-1}\,, \qquad x(t) \sim t^{-\frac{1}{2}} \,, \qquad h(t) \sim t^{-2} \, \ln{t}  \,, \qquad  g(t) \sim t^\frac{1}{2} \,, \qquad r(t) \sim t^{-3} \ln{t} \,,
          \end{align}
and the configuration is still unstable because the connection function perturbations diverge. Only if the second derivative of the potential is also zero, we find better stability properties
   \begin{align}
              \gamma_{*}(t) \sim t^{-1}\,, \qquad x(t) \sim t^{-\frac{1}{2}} \,, \qquad h(t) \sim t^{-2} \, \ln{t}\,, \qquad  g(t) \sim t^{-\frac{3}{2}}  \,, \qquad r(t) \sim t^{-3} \ln{t} \,,
         \end{align}
         and the solutions converge to general relativity.

Finally, if the model functions $f(\phi)$ and $V(\phi)$ have both an extremum at the same value of $\phi$ and the respective derivatives vanish, $f_*'= 0$, $V_*' = 0$, the background equations leave $\gamma_*(t)$ undetermined. We can integrate the first order perturbed metric and matter equations to find
   \begin{align}
             h(t) \sim t^{-2} \,, \qquad r(t) \sim t^{-3} \,.
      \end{align}
      As the first order perturbed connection equation is identically satisfied, we can invoke the second order perturbed connection equation, which is solved by
         \begin{align}
             \label{eq: radiation x_s(t) set 2}
                  x(t) &= \pm \sqrt{\frac{c_6}{\sqrt{t}}+c_7} \,.
         \end{align}
          Like in the dust matter case, a significant class of solutions with large enough initial velocities, $|\dot x_0| > \frac{|x_0|}{4 t_0}$, reach $\phi_*$ (i.e. $x=0$) with diverging speed $\dot x$, and the system experiences a singularity. Only the solutions with the integration constant $c_7=0$ converge asymptotically to $\phi_*$ without meeting a singularity. For those solutions we can solve the first order scalar field equation by
           \begin{align}
                 \gamma_*(t) &= \frac{4 V''_* t}{15 f''_*} + \frac{1}{t} - \frac{1}{12 f''_* t} + \frac{c_8}{t^\frac{3}{2}}
            \end{align}
          but the question of stability remains without a decisive answer since the connection perturbations $g(t)$ remain undetermined.  

In summary, the only of the radiation dominated configuration which is stable for perturbations around the standard general relativistic cosmological scenario is given by $f_*\neq 0$, $V_*'=V_*''=0$.
    
\subsubsection{Potential domination} \label{subsubsec:Potential domination for conn 2}
     In the era when the scalar potential dominates over the matter energy density, and we can drop $\rho(t)$ in comparison with $V({\phi_*})$ (but keep dust matter perturbations $r(t)$ with $\mathrm{w}=0$) the leading order expressions of Eqs.\ \eqref{eq: expanded cosmo set 2 FR1}, \eqref{eq: expanded cosmo set 2 FR2}, \eqref{eq: expanded cosmo set 2 ME} are solved by the standard dark energy (cosmological constant) dominated background \eqref{eq: dark energy dominant background} where $H_*=\sqrt{\tfrac{V_*}{3(1+f_*)}}$. The background part of the connection equation \eqref{eq: expanded cosmo set 2 CE} is identically zero, while the remaining leading order part of the scalar field equation \eqref{eq: expanded cosmo set 1 SE} can then be satisfied in three ways. 

    First, for an arbitrary $\phi_*$ whereby $f_*' \neq 0$, $V_*' \neq 0$, the evolution of $\gamma_*(t)$ in \eqref{eq: radiation bg gamma eq} is given by
     \begin{align}
          \label{eq: potential bg gamma sol}
              \gamma_*(t) &= c_1 e^{-3 H_* t} + \frac{2 H_*}{3} + \frac{2 V'_* }{9 H_* f'_*}\,.
       \end{align}
       Substituting the background values \eqref{eq: dark energy dominant background} and \eqref{eq: potential bg gamma sol} into the first order perturbed parts of \eqref{eq: expanded cosmo set 2} gives a system of equations to determine the behavior of perturbations, which in the leading order turn out to be
     \begin{align}
         \label{eq: potential 1p sol}
               x(t) \sim e^{-3 H_* t} \,, \qquad h(t) \sim e^{-3 H_* t} \,,\qquad  g(t) \sim t \, e^{-3 H_* t} \,, \qquad r(t) \sim e^{-3 H_* t} \,.
        \end{align}
        This regime is stable.

    Second, if $\phi_*$ is fixed by $V_*' = 0$, but $f_*' \neq 0$, the time dependence of the connection function $\gamma_*(t)$ remains as before in \eqref{eq: potential bg gamma sol}, and the first order perturbations in the leading order behave exactly as in \eqref{eq: potential 1p sol}.

    Finally, if the model functions $f(\phi)$ and $V(\phi)$ have both an extremum at the same value of $\phi$ and the respective derivatives vanish, $f_*'= 0$, $V_*' = 0$, the background equations leave $\gamma_*(t)$ undetermined. We can integrate the first order perturbed metric and matter equations to find
    \begin{align}
               h(t) \sim e^{-3 H_* t} \,, \qquad r(t) \sim e^{-3 H_* t} \,.
          \end{align}
           As the first order perturbed connection equation is identically satisfied, we can invoke the second order perturbed connection equations, which is solved by
     \begin{align}
             \label{eq: potential x_s(t) set 2}
                  x(t) &= \pm \sqrt{c_6 \, e^{-3 H_* t} + c_7} \,.
          \end{align}
Analogously to the previous cases of dust and radiation, the trajectories encounter a singularity with diverging $\dot x$ at a finite future moment if the initial values $x_0$ and $\dot x_0$ are of the opposite sign and $|\dot x_0| > 9 H_* |x_0| $. Only the class of solutions with $c_7=0$ converges to $\phi_*$. For those solutions we can solve the perturbed scalar field equation to find
           \begin{align}
                \gamma_*(t) &= c_{1} e^{- 3 H_{*} t} + \frac{2 H_{*}}{3} - \frac{H_{*}}{2 f^{\prime\prime}_*} + \frac{2 V^{\prime\prime}_*}{9 H_{*} f^{\prime\prime}_*}
           \end{align}
but the perturbations $g(t)$ remain undetermined and the stability of the system unclear.

Thus combining the analysis results of all three eras, the cosmic history with connection set 2 can be stable only if there exists $\phi_*$ which satisfies $f_*' \neq 0$, $V_*'=V_*''=0$. In such scenario, the function $\gamma(t)$ decreases in time and the solutions will converge to their respective GR behaviors. Otherwise the presence of the extra connection function diverts the cosmic evolution from the standard path. In particular, if the extrema of the model functions $f(\phi)$ and potential $V(\phi)$ coincide, a wide class of initial conditions will lead to  singular behavior of the scalar field perturbations, probably detrimental to the background dynamics.

\subsection{Connection set 3} \label{subsec:Connection set 3}
Applying the parametrization and expansion introduced in Sec.\ \ref{sec: Limit of GR} on the cosmolgical equations \eqref{eq:cosmo equations set 3} of connection set 3 yields
\begin{subequations}
\begin{align}
\label{eq: expanded cosmo set 3 FR1}
    6 (1+f_{*}) H_{*}^{2}(t) - 2 V_{*} - 2 \kappa^{2} \rho_{*}  (t) + \Big( 12 (1+f_{*}) H_{*}(t) h(t) + 6 f^{\prime}_* H_{*}^{2}(t) x(t) - 3 f^{\prime}_* \gamma_{*}(t) \dot x(t) - 2 V^{\prime}_* x(t) - 2 \kappa^{2} r(t)  \Big) &= 0 \,, \\
\label{eq: expanded cosmo set 3 FR2}
     4 (1+f_{*}) \dot H_{*}(t) + 6 (1+f_{*}) H_{*}^{2}(t) - 2 V_{*} + 2 \kappa^{2} \mathrm{w} \rho_{*}(t) + \Big( 4 f^{\prime}_* x(t) \dot H_{*}(t) + 12 (1+f_{*}) H_{*}(t) h(t) + 4 (1+f_{*}) \dot h(t) & \nonumber \\ + 6 f^{\prime}_* H_{*}^{2}(t) x(t) + 4 f^{\prime}_* H_{*}(t) \dot x(t) - f^{\prime}_* \gamma_{*}(t) \dot x(t) - 2 V^{\prime}_* x(t) + 2 \kappa^{2} \mathrm{w} r(t) \Big)  &= 0 \,, \\
\label{eq: expanded cosmo set 3 CE}
     15 f^{\prime}_* H_{*}(t) \gamma_{*}(t) \dot x(t) + 3 f^{\prime}_* \gamma_{*}(t) \ddot x(t) + 6 f^{\prime}_* \dot \gamma_{*}(t) \dot x(t)
     + \Big(15 f^{\prime\prime}_* H_{*}(t) \gamma_{*}(t) x(t) \dot x(t) + 3 f^{\prime\prime}_* \gamma_{*}(t) x(t) \ddot x(t) & \nonumber \\  + 3 f^{\prime\prime}_* \gamma_{*}(t) \left(\dot x(t)\right)^{2} + 6 f^{\prime\prime}_* x(t) \dot \gamma_{*}(t) \dot x(t) + 15 f^{\prime}_* H_{*}(t) g(t) \dot x(t) + 15 f^{\prime}_* \gamma_{*}(t) h(t) \dot x(t) + 3 f^{\prime}_* g(t) \ddot x(t) + 6 f^{\prime}_* \dot g(t) \dot x(t) \Big) &=0 \,, \\
\label{eq: expanded cosmo set 3 SE}
      3 f^{\prime}_* \left( \dot \gamma_{*}(t) - 2 H_{*}^{2}(t) + 3 H_{*}(t) \gamma_{*}(t) \right) - 2 V^{\prime}_* 
     - \Big( 2 V^{\prime\prime}_* x(t) + 6 f^{\prime\prime}_* H_{*}^{2}(t) x(t) - 9 f^{\prime\prime}_* H_{*}(t) \gamma_{*}(t) x(t) & \nonumber \\ - 3 f^{\prime\prime}_* x(t) \dot \gamma_{*}(t) - 9 f^{\prime}_* H_{*}(t) g(t) + 12 f^{\prime}_* H_{*}(t) h(t) - 9 f^{\prime}_* \gamma_{*}(t) h(t) - 3 f^{\prime}_* \dot g(t) + 6 H_{*}(t) \dot x(t) + 2 \ddot x(t) \Big) &=0 \,, \\
\label{eq: expanded cosmo set 3 ME}
     \dot \rho_{*}(t) + 3 (1+\mathrm{w}) H_{*}(t) \rho_{*}(t) + \Big( 3 \mathrm{w} H_{*}(t) r(t) + 3 \mathrm{w} \rho_{*}(t) h(t) + 3 H_{*}(t) r(t) + 3 \rho_{*}(t) h(t) + \dot r(t) \Big) & = 0 \,.
\end{align}
\end{subequations}
As before, we will consider the stability of the equations in the case of different matter types separately. The calculations are rather similar to the dust matter case in the previous section, and we will just present the main results here.

\subsubsection{Dust matter domination} \label{subsubsec:Dust matter domination for conn 3}
In the regime of dust matter domination we take $\mathrm{w}=0$ and assume $V_*$ is negligible in comparison with $\rho(t)$. At the background level the metric and matter equations \eqref{eq: expanded cosmo set 3 FR1}, \eqref{eq: expanded cosmo set 3 FR2}, \eqref{eq: expanded cosmo set 3 ME} coincide with the GR cosmological equations \eqref{eq: GR Friedmann equations} and are solved by the standard background evolution \eqref{eq: dust dominant background}. The background part of the connection equation \eqref{eq: expanded cosmo set 3 CE} is identically zero, while in solving the scalar field equation \eqref{eq: expanded cosmo set 3 SE} at the background level, 
    \begin{align}
        3 f'_* \left( \dot{\bar\gamma}_* + \frac{2 \bar\gamma_*}{t} - \frac{8}{9 t^2} \right) - 2 V'_* &=0 \,,
     \end{align}
there are three options. 

First, if $f'_* \neq 0$ and $V'_* \neq 0$, then to the leading order the background connection function obeys
   \begin{align}
        \bar{\gamma}_* &= \frac{2 V'_* t}{9 f'_*} + \frac{8}{9t} + \frac{c_1}{t^2} \,.
     \end{align}
Substituting that into the first order small equations we find that the perturbations evolve as
     \begin{align}
         h &\sim t^{-2} \,, \qquad r \sim t^{-3} \,, \qquad x\sim t^{-\frac{13}{3}} \,, \qquad g \sim t^0
     \end{align}
and the configuration is marginally stable. Although $\bar{\gamma}_*$ increases in time, its effects are not visible, since $\dot{\phi}$ decreases and in the Friedmann equation the combined term evolves as $\bar{\gamma}\dot{\phi} \sim t^{-\frac{13}{3}}$.

Second, if $f'_* \neq 0$ but $V'_* = 0$, then to the leading order
  \begin{align}
        \bar{\gamma}_* &\sim t^{-1} \,,
    \end{align}
and the first order small equations give
    \begin{align}
         h &\sim t^{-\frac{4}{3}} \,, \qquad r \sim t^{-\frac{7}{3}} \,,  \qquad x\sim t^{-\frac{1}{3}} \,,  \qquad   g \sim t^{\frac{2}{3}} \,.
    \end{align}
This situation is unstable due to the growth of connection function perturbations $g(t)$, which at some moment would spoil the assumption that all perturbations are small. However, if in addition also $V_*'' = 0$, then the leading order solution becomes
    \begin{align}
         h &\sim t^{-\frac{4}{3}} \,, \qquad r \sim t^{-\frac{7}{3}} \,,  \qquad x\sim t^{-\frac{1}{3}} \,,  \qquad   g \sim t^{-\frac{4}{3}} \,
    \end{align}
and the regime can be considered to be stable instead.

Thirdly, if the model functions allow a scalar field value where simultaneously $f'_* = 0$ and $V'_* = 0$, then the connection field equation is automatically satisfied at the leading as well as first perturbative order and we can not determine $\bar{\gamma}_*$ from there. The other first order small equations give
       \begin{align}
               h &\sim t^{-2} \,, \qquad r \sim t^{-3} \,.
       \end{align}
To find $\bar{\gamma}_*(t)$ and $x(t)$ we have the first perturbation of the scalar field equation \eqref{eq: expanded cosmo set 3 SE} and second perturbation of the connection equation \eqref{eq: expanded cosmo set 3 CE} at our disposal. 
It is a coupled system for $\bar{\gamma}_*(t)$ and $x(t)$ and straightforward integration seems difficult. However, incidentally the connection equation \eqref{eq: expanded cosmo set 3 CE} with $h(t)$ and $r(t)$ and substituted in, can be solved by
\begin{align}
    \bar{\gamma}_* &= \frac{c_9}{t^{\frac{5}{3}}\sqrt{|x||\dot{x}|}} \,.  
\end{align}
We can substitute that expression into the remaining equation \eqref{eq: expanded cosmo set 3 SE} which yields
\begin{align}
\label{eq: dust x_s(t) set 3 equation}
    \ddot{x} &= -\frac{\dot{x} \left(3 c_{9} f^{\prime\prime}_* \left(3 t \dot{x} - 2 x\right) + 4 t^{\frac{2}{3}} \left(3 V^{\prime\prime}_* t^{2} x + 4 f^{\prime\prime}_* x + 6 t \dot{x}\right) \sqrt{\left|{x}\right|\left|{\dot{x}}\right|}\right)}{3 t \left(3 c_{9} f^{\prime\prime}_* x + 4 t^{\frac{5}{3}} \dot{x} \sqrt{\left|{x}\right|\left|{\dot{x}}\right|}\right)} \,.
\end{align}
This is by far a more complicated equation than the corresponding Eq.\ \eqref{eq: dust x_s(t) set 2 equation} which emerged for connection set 2. It is rather hard to solve it analytically, but we can still discern the main characteristics by studying the equation in different limits, complemented by sample phase portraits on Fig.\ \ref{fig: phase portraits}. Since Eq.\ \eqref{eq: dust x_s(t) set 3 equation} depends explicitly on time $t$, the actual phase space is three dimensional. Furthermore, it depends on the values of the parameters $V_*''$, $f_*''$, and $c_9$. However, to gain a glimpse of the principal features of the dynamics, Figs. \ref{fig: phase portrait set3b dust} and \ref{fig: phase portrait set3c dust} present two illustrative phase portraits for fixed values of parameters and time, that capture the qualitative behavior of available cases. 

In the limit $x\to 0$ we can expand \eqref{eq: dust x_s(t) set 3 equation}  to get 
\begin{align}
    \ddot{x} &= - \frac{3 c_9 f_*'' \sqrt{|\dot{x}|}}{4 t^{\frac{5}{3}} \sqrt{x}} + \mathcal{O}(x^0) \,,
\end{align}
and see that the force diverges. Hence the solutions experience a singularity, depicted by a red line between the regions $(I)$ and $(II)$ on the plots \ref{fig: phase portrait set3b dust} and \ref{fig: phase portrait set3c dust}. The direction of the force (accelerating or decelerating) depends on the signs of the parameters $c_9$ and $f_*''$. Next, in the limit $\dot{x} \to 0$ we get from \eqref{eq: dust x_s(t) set 3 equation} 
\begin{align}
    \ddot{x} &= \frac{2 \dot{x}}{3t} + \mathcal{O}(\dot{x}^{\frac{3}{2}}) \,,
\end{align}
which tells that the standstill state $\dot{x}=0$ is unstable and any small deviation from it will meet a force (``anti-friction'') pushing the solutions away. This is marked by a red line between the regions $(I)$ and ($III$) on the figures \ref{fig: phase portrait set3b dust} and \ref{fig: phase portrait set3c dust}, and is a common feature for all values of parameters. Thus, contrary to the set 2 case on \ref{fig: phase portrait set2 dust} the solutions can not come stabilize at some value of the scalar field.
Finally, we notice that the expression \eqref{eq: dust x_s(t) set 3 equation} has also another string of singularities at 
\begin{align}
x_s&= -\frac{16 t^{\frac{10}{3}} \dot{x}^3}{9 c_9^2 (f_*'')^2} \mathrm{sign}(c_9 f_*'')
\end{align}
where the denominator on the r.h.s.\ vanishes. This is shown as the red line between the regions $(II)$ and ($III$) on the plots. That string of singularities can act as a source or sink for the neighbouring trajectories, depending on the sign of the numerator in Eq.\ \eqref{eq: dust x_s(t) set 3 equation}. The point where the numerator vanishes, and this singular curve switches between repeller and attractor behaviors is marked by an enlarged dot on the plots.

In short, for the case on plot \ref{fig: phase portrait set3b dust} the available classes of solutions can be summarized as follows. First, if the initial conditions $x_0$ and $\dot{x}_0$ are of the same sign, i.e.\ regions $(I)$, the solutions have either started from an initial singularity at $x=0$ or from an unstable state at $\dot{x}=0$, and consistently flow away from the value $\phi_*$. Alternatively, if $x_0$ and $\dot{x}_0$ are of opposite signs, either in region $(II)$ with $|x|<|x_s|$ or region $(III)$ with $|x|>|x_s|$, the solutions inevitably crash into a singularity where $\ddot{x}$ diverges. Thus, none of the solutions can actually manage to reach the point $\phi_*$ and stabilize there ($x=0$, $\dot{x}=0$). 

For the case on plot \ref{fig: phase portrait set3c dust} the available classes of solutions can be summarized in a similar manner. First, if the initial conditions $x_0$ and $\dot{x}_0$ are of the opposite sign, i.e.\ regions $(I)$, the solutions have either started from the unstable state at $\dot{x}=0$ and flow towards the singularity of $|x|\to 0$, $|\dot{x}|\to \infty$. Alternatively, if $x_0$ and $\dot{x}_0$ are of the same sign, either in region $(II)$ with $|x|<|x_s|$ or region $(III)$ with $|x|>|x_s|$, the solutions inevitably crash into another singularity where $\ddot{x}$ diverges. Hence, none of the solutions can actually manage to reach the point $\phi_*$ and stabilize there.
Therefore, we don't need to consider the scenario of $f_*'=0$, $V_*'=0$ any further (estimating $\bar{\gamma}(t)$, $g(t)$), but just conclude that this regime is unstable.

In summary, the dust matter dominated regime for the connection set 3 can be stable if $f_*' \neq 0$ and $V_*'=V_*''=0$. However, if a model allows a value $\phi_*$ where simultaneously $f_*' =0$ and $V_*'=0$, then a large class of solutions will likely face a singularity in finite time.

\subsubsection{Radiation domination} \label{subsubsec:Radiation domination for conn 3}
In the radiation domination regime we take the barotropic index $\mathrm{w}=\tfrac{1}{3}$ and assume $V_*$ is negligible in comparison with $\rho(t)$. At the background level the metric and matter equations \eqref{eq: expanded cosmo set 3 FR1}, \eqref{eq: expanded cosmo set 3 FR2}, \eqref{eq: expanded cosmo set 3 ME} coincide with the GR cosmological equations \eqref{eq: GR Friedmann equations} and are solved by the standard background evolution \eqref{eq: radiation dominant background}. The background part of the connection equation \eqref{eq: expanded cosmo set 3 CE} is identically zero, while in solving the scalar field equation \eqref{eq: expanded cosmo set 3 SE} at the background level, 
     \begin{align}
         3 f'_* \left( \dot{\bar\gamma}_* + \frac{3 \bar\gamma_*}{2t} - \frac{1}{2 t^2} \right) - 2 V'_* &=0  \,,
    \end{align}
there are again three possible cases.
 
First, if $\phi_*$ is arbitrary in the sense that $f_{*}'\neq 0$ and $V_{*}'\neq 0$, then the connection background solution is given by
 \begin{align}
        \bar{\gamma}_*(t) &= \frac{4 V_*'t}{15 f_*'} +\frac{1}{t}+\frac{c_1}{t^{\frac{3}{2}}} \,, 
\end{align}
while taking into account only the leading orders the perturbations evolve as
\begin{align}
        x(t) \sim t^{-\frac{7}{2}} \,, \qquad h(t) \sim t^{-2} \,, \qquad r(t) \sim t^{-3}\,,\qquad g(t) \sim t^0 \,.
\end{align}
This regime is only marginally stable. 

Second, if $f_{*}'\neq 0$ but $V_{*}'= 0$, then 
 \begin{align}
        \bar{\gamma}_*(t) &\sim t^{-1} \,,
\end{align}
and the perturbations solve the equations in the leading order as
 \begin{align}
        x(t) \sim t^{\frac{1}{2}} \,, \qquad h(t) \sim t^{-2} \,, \qquad r(t) \sim t^{-3} \,,\qquad g(t) \sim t^{\frac{3}{2}} \,,
    \end{align}
which is unstable. If in addition also $V_{*}''=0 $ then 
\begin{align}
        x(t) \sim t^{\frac{1}{2}} \,, \qquad h(t) \sim t^{-2} \,, \qquad r(t) \sim t^{-3} \,,\qquad g(t) \sim t^{-\frac{1}{2}} \,,
\end{align}
but the regime is still unstable as the scalar field perturbations are not under control.

Third, if there exists such $\phi_*$ that $f_{*}'= 0$ and $V_{*}'= 0$ then we obtain
\begin{align}
       \qquad h(t) \sim t^{-2} \,, \qquad r(t) \sim t^{-3}\,.
\end{align}
Like in the dust case, the first order perturbation of the connection equation \eqref{eq: expanded cosmo set 3 CE} is identically satisfied. To proceed we can turn to the second order perturbation of the connection equation, which can be solved by
\begin{align}
    \bar{\gamma}_* &= \frac{c_9}{t^{\frac{5}{4}}\sqrt{|x||\dot{x}|}} \,. 
\end{align}
Substituting that into the first order scalar field equation \eqref{eq: expanded cosmo set 3 SE} yields again a highly nonlinear equation
\begin{align}
\label{eq: radiation x_s(t) set 3 equation} 
    \ddot{x}(t)=-\frac{\dot{x}\left(3c_{9}f_{*}^{\prime\prime}(2t\dot{x}-x)+2t^\frac{1}{4}(3f_{*}^{\prime\prime}x+4V_{*}^{\prime\prime}t^{2}x+6t\dot{x})\sqrt{|x||\dot{x}|}\right)}{2t\left(3c_{9}f_{*}^{\prime\prime}+4t^\frac{5}{4}\dot{x}\sqrt{|x||\dot{x}|}\right)}\,.
\end{align}
This is structurally analogous to Eq.\ \eqref{eq: dust x_s(t) set 3 equation} from the dust matter dominated case with only slightly differing numerical factors and one power of $t$. Thus the phase portraits are qualitatively similar to Figs.\ \ref{fig: phase portrait set3b dust}, \ref{fig: phase portrait set3c dust}, and the so are also the results of the analysis about the solutions. The conclusion is that the dynamics around $f_{*}'= 0$ and $V_{*}'= 0$ is unstable, and can lead to a singular behavior.

In summary, we see that the radiation dominated regime is unstable, even the case of $f_{*}'\neq 0$ but $V_{*}'= V_{*}''=0$. Again, if a model allows a value $\phi_*$ where simultaneously $f_*' =0$ and $V_*'=0$, then a large class of solutions will likely face a singularity in finite time.

\subsubsection{Dark energy domination} 
\label{subsubsec:Dark energy domination for conn 3}
In the dark energy domination regime we assume $V_*$ is much larger than $\rho(t)$, but keep the dust matter perturbations $r(t)$ with $\mathrm{w}=0$. At the background level the metric and matter equations \eqref{eq: expanded cosmo set 3 FR1}, \eqref{eq: expanded cosmo set 3 FR2}, \eqref{eq: expanded cosmo set 3 ME} coincide with the GR cosmological equations \eqref{eq: GR Friedmann equations} and are solved by the standard background evolution \eqref{eq: dark energy dominant background} where $H_* = \sqrt{\tfrac{V_*}{3(1+f_*)}}$. The background part of the connection equation \eqref{eq: expanded cosmo set 3 CE} is identically zero, while solving the scalar field equation \eqref{eq: expanded cosmo set 3 SE} at the background level, 
     \begin{align}
     \label{eq: SE background set 3}
         3 f'_* \left( \dot{\bar\gamma}_* + 3 H_* \bar\gamma_* - 2 H_*^2 \right) - 2 V'_* &=0  \,,
     \end{align}
again allows three possible cases.

First, if $f_{*}'\neq 0$ and $V_{*}'\neq 0$ then Eq.\ \eqref{eq: SE background set 3} is solved by
\begin{align}
    \bar{\gamma}_*(t) &=  c_1 e^{-3H_{*}t} + \frac{2H_*}{3} + \frac{2 V_*'}{9 H_* f_*'} 
\end{align}
and up to the leading order the perturbations evolve as
 \begin{align}
        x(t) \sim e^{-5H_{*}t} \,, \qquad h(t) \sim e^{-3H_{*}t}  \,, \qquad r(t) \sim e^{-3H_{*}t}\,,\qquad g(t) \sim te^{-3H_{*}t} 
    \end{align}
which is a stable situation.

Second, $f_{*}'\neq 0$ but $V_{*}'= 0$ some terms in the equations drop out, but the leading order behavior of $\bar\gamma_*$ and the perturbations do not change compared to the above. Similarly, taking also $V_*''=0$ retains the picture of convergence. 

Thirdly, if $f_{*}'= 0$ and $V_{*}'= 0$ we can solve the equations by the same means as in the analogous case of set 3 dust matter domination, to find that the leading behavior is
\begin{align}
    r(t) &\sim e^{-3H_{*}t} \,, \qquad h(t) \sim e^{-3H_{*}t}\,.   
\end{align}
% , \qquad x(t) \sim \ljtext{e^{-5H_{*}t}} \,, \lptext{\sim t^{\frac{1}{2}}e^{\frac{-5H_{*}t}{2}}}
To obtain $\gamma_{*}(t)$, Since the background connection equation \eqref{eq: expanded cosmo set 3 CE} is identically satisfied, we need to go the second-order small connection equation
\begin{align} \label{DE_Gamma_conn3}
    \gamma_*(t)=\frac{c_{9}}{e^{\frac{5 H_{*}t}{2}}\sqrt{|x||\dot{x}|}}
\end{align}
Using the above equation \eqref{DE_Gamma_conn3} into the 1st order small scalar field equation \eqref{eq: expanded cosmo set 3 SE} we obtained again a highly nonlinear differential equation
\begin{align}
    \ddot{x}(t)=-\frac{\dot{x}\left(3c_{9}f_{*}^{\prime\prime}(\dot{x}-H_*x)+4 e^{\frac{5 H_* t}{2}}(3f_{*}^{\prime\prime} H_*^2 x+V_{*}^{\prime\prime}x+3 H_*\dot{x})\sqrt{|x||\dot{x}|}\right)}{3c_{9}f_{*}^{\prime\prime}x+4 e^{\frac{5 H_* t}{2}} \dot{x}\sqrt{|x||\dot{x}|}}\,.
\end{align}
Apart from the details of the explicit time factors, this equation is analogous to Eqs.\ \eqref{eq: dust x_s(t) set 3 equation} and \eqref{eq: radiation x_s(t) set 3 equation} we met in the dust and radiation cases. A closer analysis reveals the same key features of dynamical behavior. It is not possible for the system to relax at $\phi_*$ where $f_{*}'= 0$ and $V_{*}'= 0$, but the solutions either flow away from this value, or the the system encounters a singularity where $\ddot{x}$ diverges and the approximation of small perturbations breaks down.

In summary, the potential dominated regime is stable if $f_{*}'\neq 0$. On the other hand, for $f_{*}'= 0$ and $V_{*}'= 0$ large classes of solutions meet a finite singularity. In the end, combining the analysis results of all three eras, it seems a cosmic history from radiation to dust to dark energy domination can not be realized in a stable manner in the vicinity of some fixed scalar field value $\phi_*$. 
%Radiation era is stable around $f_*' = 0$, $V_*'=0$, but that becomes unstable in the dust domination epoch. 
Dust matter and potential domination eras are stable around $f_*' \neq 0$, $V_*'=V_*''=0$, but that is not stable in the radiation domination epoch.
 
\section{Discussion} \label{sec:conclusion}

In this paper we have explored the cosmological implications of alternative FLRW connections that become available in symmetric teleparallel geometry, focusing on the analogue of scalar-tensor gravity where a scalar field is nonminimally coupled to the nonmetricity scalar in the action \eqref{Action}. A demand that the independent connection with zero curvature and zero torsion obeys the symmetries of spatial homogneneity and isotropy was recently shown to yield three sets of connections which involve an extra free function of time \cite{Hohmann:2021ast,DAmbrosio:2021pnd}. In the first set this extra function actually drops out of the cosmological equations, and the system reduces to the one already known to correspond to a trivial connection and also to coincide with the scalar-torsion case in metric teleparallelism, which has been studied a lot. In the two other classes, however, the extra function is present in the equations, and could be interpreted as in instance of an extra degree of freedom that has been notoriously difficult to pinpoint down in extended teleparallel gravities before.

In section \ref{sec:scalartensorcosmology} we presented the cosmological field equations arising for these connections, and confirmed that the new function does indeed increase the number of independent phase space dimensions by one, thus it is not a constraint in disguise. We also observed from the Friedmann equations that the new function can not itself take the role of dark energy or dark matter, rather it behaves as stiff fluid or spatial curvature in connection sets 2 or 3, respectively. Furthermore, in the Friedmann equations the extra function only appears if the nonminimal coupling $\mathcal{A}(\Phi)$ between the scalar field and nonmetricity is not constant and the time derivative of the scalar field is not zero. Hence it has no effect in the case of minimally coupled fields or when the dynamics of the scalar field has stopped, i.e.\ in the cases when the model is equivalent with GR. In addition we also found that the extra function drastically modifies the scalar field dynamics, since the connection equation can be viewed as a dynamical equation for the scalar field, albeit without any contribution from the kinetic coupling $\mathcal{B}(\Phi)$ or scalar potential $\mathcal{V}(\Phi)$, see Eqns.\ \eqref{eq: cosmology eq set 2 Phi_tt} and \eqref{eq: cosmology eq set 3 Phi_tt} which look quite puzzling. Moreover, these equations also pose a danger to throw the scalar field into a singular state (infinite $\ddot{\Phi}$) when the nonminimal coupling function has an extremum ($d\mathcal{A}/d\Phi=0$), again a rather problematic feature, although possible to mitigate with monotonic functions. It is hard to declare in full generality for arbitrary model functions, but in the cosmological equations it is not obvious how the extra function in the connection could offer new options to generate dark energy, besides the well known regime of slowly evolving scalar field with a positive potential acting similar to a cosmological constant.

Owing to the strange features the alternative connections introduce into the system, we proceeded in Sec.\ \ref{sec: Limit of GR} and \ref{sec:Stability of standard cosmological regimes} to study whether and under which conditions the standard cosmological eras are stable in the model, i.e.\ that the succession of eras is not disturbed by other factors than the densities of different matter components decreasing at different rate as the universe expands. In the scalar-tensor context the scalar field value should also not change too much from early on, in order to satisfy the observational constraints. The vague expectation is that the current tensions in the data could be eventually explained by suitable convergence processes or oscillations around the standard scenario. Thus, we expanded the scalar-nonmetricity FLRW equations around the $\Lambda$CDM background with the radiation domination, dust matter domination, and potential domination assumptions, solved them explicitly and determined the asymptotic behavior of all quantities. In a stable situation, all perturbations should decrease in time. In essence the question is whether the ``attractor mechanism'' that is known in Riemannian scalar-tensor cosmology \cite{Damour:1992kf,Damour:1993id,Serna:1995pi,Mimoso:1998dn,Santiago:1998ae,Serna:2002fj,Jarv:2010zc,Jarv:2010xm,Jarv:2011sm,Jarv:2015kga,Dutta:2020uha} has an analogue in the symmetric teleparallel counterpart. 

For the connection set 1 this investigation can be carried over from the metric teleparallel case which has the same equations \cite{Jarv:2015odu}. 
In the parametrization where $\mathcal{A}=1+f(\phi)$, $\mathcal{B}=1$, $\mathcal{V}=V(\phi)$, all three eras are stable if the model functions allow a value $\phi_*$ which corresponds to a simultaneous minimum of the gravitational coupling function ($f'_*=0$, $f''_*>0$) and a minimum or at least an inflection point of the potential ($V_*'=0, V''_*\geq 0$).  It is notable, that while in the Riemannian scalar-tensor case during the radiation era the scalar field stabilizes to an arbitrary value which can be very different from the value it will be drawn to in the matter and potential domination eras (thus possibly causing a large drift in the gravitational constant at the beginning of matter domination), in the symmetric teleparallel case of connection set 1 the scalar field value would then remain stable already since the radiation era. 

For the alternative connection sets our results are new and can be summarized as follows. The cosmic history with connection set 2 can be stable through all three eras only if there exists $\phi_*$ which satisfies $f_*' \neq 0$, $V_*'=V_*''=0$. In such scenario, the extra connection function decreases in time and the solutions converge to their respective GR behaviors. Otherwise the presence of the extra connection function diverts the cosmic evolution from the standard path. In contrast, for connection set 3, the dust matter and potential domination eras are stable around $f_*' \neq 0$, $V_*'=V_*''=0$, but there is no stable configuration for the radiation domination epoch. At face value these stability properties look fine, although in comparison with the connection set 1 it is weird why the minima of the coupling function or potential do not figure in the stability conditions. A further reflection, however, realizes a deeper problem. The gradient of the potential that would normally act as a force term in the evolution equation, has no role in determining the scalar field dynamics, see again Eqns.\ \eqref{eq: cosmology eq set 2 Phi_tt} and \eqref{eq: cosmology eq set 3 Phi_tt}. The scalar field experiences just friction and anti-friction type of influence, and only rather randomly can end up at the point of stability. Or the scalar field can experience a singularity, if the extrema of the model functions coincide, $f'(\phi_s)=V'(\phi_s)$. This feature is illustrated on the phase portraits of Fig.\ \ref{fig: phase portraits}, and probably means that the other quantities become singular as well.

In conclusion, the alternative FLRW connections can not be deemed outright pathological and do not make the universe definitely unstable, but they have a very strange and possibly dangerous influence on the scalar field dynamics nevertheless. Further studies are needed to understand whether this influence is overwhelmingly harmful, or could be useful in describing some phenomena in the end. It might happen, that although compatible with FLRW symmetry, the alternative connections get eventually ruled out, when we learn how to correctly implement the boundary term in the actions of extended teleparallel gravities to fix the connection.

\subsection*{Acknowledgements}This work was supported by the European Regional Development Fund through the Center of Excellence TK133 ``The Dark Side of the Universe,'' and the Estonian Research Council grants PRG356 ``Gauge Gravity'' and TK202 ``Foundations of the Universe.'' We are grateful to the referee at Phys.\ Rev.\ D for insightful comments. %L.P. has been supported by the European Regional Development Fund CoE program TK133, “ The Dark Side of the Universe.”

  % \bibliographystyle{utphys.bst}    % VERY NICE STYLE
  % \bibliography{references}
  
\providecommand{\href}[2]{#2}\begingroup\raggedright\endgroup

\end{document}